\documentclass[a4paper,11pt]{article}
\usepackage{amsmath,amssymb,ascmac,enumerate,booktabs,amsthm}
\usepackage[dvips]{graphicx}
\usepackage[utf8]{inputenc}
\usepackage{lscape}
\usepackage{url}
\usepackage[dvipdfmx]{hyperref}
\usepackage{braket}
\usepackage{color}

\newcommand{\1}{\mbox{1}\hspace{-0.25em}\mbox{l}}

\usepackage[dvipdfmx]{hyperref}

\makeatletter
 
  \@addtoreset{equation}{section}
\makeatother

\begin{document}

\begin{titlepage}
\begin{flushright}
DIAS-STP-24-30 \\
\today
\end{flushright}
\vspace{0.5cm}
\begin{center}
{\Large \bf
Genus 2 Superstring Chiral Measure
\\
From The $3$-Dimensional Gelca-Hamilton TQFT
}

\lineskip .75em
\vskip 2.5cm
{\large  Saki Koizumi \footnote{saki@stp.dias.ie}}
\vskip 2.5em
 {\normalsize\it School of Theoretical Physics,
Dublin Institute for Advanced Studies,\\
$10$ Burlington Road, Dublin $4$, Ireland}
\vskip 3.0em


\end{center}


\begin{abstract}
In the path integral formulation of the superstring, the chiral measure acquires a phase under the modular transformation of a Riemann surface.
This motivated the use of anomaly inflow to define the superstring chiral measure by a path integral formalism of a modular invariant $3$-dimensional theory.
A Gelca-Hamilton topological field theory (TQFT) is one of the Atiyah's TQFT on a $3$-dimensional extended manifold with the boundary Jacobi variety of a Riemann surface, whose Hilbert space is spanned by the theta series.
We show that genus $g\leq 2$ superstring chiral measure in the path integral can be obtained by the path integral of the Gelca-Hamilton TQFT on some $3$-dimensional bulk extended manifolds.
The modular transformation of the superstring chiral measure can be understood as the action of the extended mapping class group on the bulk $3$-dimensional extended manifolds.

\end{abstract}

\end{titlepage}
\baselineskip=0.5cm

\tableofcontents

\section{Introduction}
When we consider the path integral of superstring theories by using the NSR formalism, the worldsheet can be regarded as a spin curve (Riemann surface with a spin structure).
The genus $1$ and genus $2$ superstring chiral measures can be expressed using the characteristic theta functions on the Jacobi variety of the spin curve.
The mapping class group on a curve induces modular transformations on the Jacobi variety.
The superstring chiral measure should be determined in a modular invariant way.
For the genus $1$ case, the superstring chiral measure is given by the product and summation of the characteristic theta functions, by the summation of the spin sectors (GSO projection).
In the genus $2$ case, it also takes the form of the linear summation of the products of characteristic theta functions \cite{Hok-Pho,DPho1,DPho2,DPho3,DPho4,DPho-lec}.

The path integral of the superstring on the spin curve is formulated via the super Riemann surface.
Since there is a one-to-one correspondence between a spin curve and a super Riemann surface \cite{Bat79}, the worldsheet can be identified as a super Riemann surface.
Then, the modular invariance is related to super-diffeomorphism invariance.
In the universal family of super Riemann surfaces, each fiber is a super Riemann surface, identified under large super-diffeomorphisms, and transforming under infinitesimal super-diffeomorphisms.
By integrating out the Grassmann odd part of the supermoduli space, we can obtain the genus $g\leq 2$ superstring chiral measure \cite{Hok-Pho,DPho1},
however, in the case of higher genus super Riemann surfaces, this does not provide the superstring chiral measure \cite{DW-SM,Gr10}.

The papers \cite{DPho-HL1,DPho-HL2} imposed constraint conditions on the form of the genus $3$ superstring chiral measure, however, found that there are no solutions that satisfy those conditions.
In the paper \cite{3-loop}, weaker conditions were considered, and solutions which coincide with the genus $1$ and genus $2$ superstring chiral measures were found.
The paper \cite{Gr} generalized the forms of genus $g\leq 3$ superstring chiral measure for genus $4$ and genus $5$, by assuming the conditions of \cite{DPho-HL1,DPho-HL2}.
This result was then considered from the alternate perspective of moduli space by \cite{GF12}.
The ansatz to obtain vanishing cosmological constant in the genus $5$ superstring chiral measure was found by \cite{Man08,Gr-Man11}.
An alternative genus $5$ ansatz was considered in \cite{OCRD-g5,genus5}, giving differing results from \cite{Man08,Gr-Man11}.

In this paper, we will study the chiral measure of the superstring path integral from the viewpoint of $3$-dimensional topological quantum field theory (TQFT), which then gives another interpretation of the chiral measure.
According to the papers \cite{Cal-Har,DT-KW}, the anomalous phase of an axion string coupled to a Dirac fermion is canceled by the bulk $3$-dimensional monopole background theory, called the anomaly inflow.
Although the superstring has no anomaly for infinitesimal super-diffeomorphisms, its chiral measure is changed by modular transformations.
Thus, we are motivated to treat the superstring chiral measure as an anomalous partition function, use the idea of anomaly inflow \cite{WY}, and identify it as the path integral of a $3$-dimensional non-anomalous theory.
In the genus $1$ and genus $2$ cases, the superstring chiral measure is obtained as a linear combination of products of the characteristic theta functions, which are sections of line bundles on the Jacobi variety of a curve.
The Jacobi variety can be identified as the extended surface, and then its bulk is an extended $3$-dimensional manifold. This manifold is a pairing of the $3$-dimensional bulk of a curve, and the framing of the $3$-dimensional manifold.
On an extended $3$-dimensional manifold, the $\mathbb{Z}$-extension of the mapping class group acts, called the extended mapping class group.
The restriction of the extended mapping class acts on the boundary Jacobi variety, and it induces a transformation of the characteristic theta functions.
However, it is difficult to obtain this induced mapping directly.

On the other hand, by identifying the Jacobi variety as a symplectic space, we can perform geometric quantization, to obtain a Hilbert space which is spanned by the theta series \cite{knot-theta}.
A representation of the extended mapping class group on the theta series is obtained by \cite{knot-theta}.
Since the theta series is a section on a line bundle on the Jacobi variety, the modular transformation on the Jacobi variety induces a mapping on the theta series.
This action of the modular group on the theta series is equivalent to the representation of the extended mapping class group on the theta series up to the multiplication of the eighth root of the unity.
We expand the characteristic theta function by a linear combination of the theta series and obtain the representation of the extended mapping class group on the characteristic theta function.
Then, we expand the genus $g\leq 2$ superstring chiral measure by the theta series, and find that the action of the extended mapping class group is equivalent to the modular transformation on the superstring chiral measure.
The path integral of the Gelca-Hamilton TQFT on a bulk $3$-dimensional extended manifold is valued in the Hilbert space on the Jacobi variety \cite{knot-theta}, where the Gelca-Hamiton TQFT is one of the Atiyah's TQFT defined in the paper \cite{knot-theta}.
Then, we obtain the genus $1$ and genus $2$ superstring chiral measures by the path integral of the Gelca-Hamilton TQFT on sets of $3$-dimensional bulk extended manifolds.
The change of the genus $2$ superstring chiral measure by a modular transformation is represented as the change of the $3$-dimensional extended manifold by adding the mapping cylinder.

The rest of this paper is organized as follows.
In section 2, we explain how the mapping class group action on a Riemann surface induces the modular transformation on the Jacobi variety.
We also determine the modular transformation of the spin structure on a Riemann surface.
We explain the characteristic theta function as a section on a line bundle on the Jacobi variety.
In section 3, we consider the superstring chiral measure in NSR formalism.
We explain that the superstring worldsheet can be regarded as a spin curve, and it is also possible to treat it as a super Riemann surface.
We explain that in the genus $g\leq 3$ case, the modular invariance of the superstring on a spin curve corresponds to the super-diffeomorphism invariance of the superstring on a super Riemann surface.
After that, we introduce the genus $1$ and genus $2$ superstring chiral measure.
In section 4, we introduce the extended $3$-dimensional manifold whose boundary is the Jacobi variety.
Then, we explain how the $\mathbb{Z}$-extension of the mapping class group (extended mapping class group) acts on the extended $3$-dimensional manifold.
We consider the geometric quantization of the Jacobi variety by identifying the Jacobi variety as a symplectic space, and obtain a Hilbert space which is spanned by the theta series.
We will review the Gelca-Hamilton TQFT \cite{knot-theta}.
In section 5, we rewrite the superstring chiral measure by the theta series and obtain it by a path integral of the bulk $3$-dimensional Gelca-Hamilton TQFT.
We confirm that modular transformation can be regarded as the extended mapping class group action on the superstring chiral measure in the Gelca-Hamilton formalism.
In Appendix A, we explain the Abelian variety, which is important to explain the Jacobi variety.
In Appendix B, we explain the symplectic automorphism action on a quadratic form.
We will explain how the Arf-invariant of quadratic forms determines the orbit of the symplectic group.
In Appendix C, we show an isomorphism of two symplectic spaces, which is necessary to show that modular transformations are induced by the mapping class group.

\section{Jacobi Variety and Characteristic Theta Function}
To study the partition function of the superstring theory, we begin with explaining the Jacobi variety of a spin curve and the modular transformation on the Jacobi variety.
We will generalize the modular transformation on a spin structure.
We will then introduce the characteristic theta functions as sections of line bundles on the Jacobi variety.
It is important because the superstring chiral measures for genus $1$ and $2$ cases can be written by product and summation of the characteristic theta functions \cite{Hok-Pho,DPho1,DPho2,DPho3,DPho4}.

\subsection{Jacobi variety and Modular Transformation}
In this subsection, we will introduce the Jacobi variety for Riemann surfaces.
We will show that the modular transformation is induced by the mapping class group action on a Riemann surface.

Let us consider a Riemann surface $\Sigma_g$ of genus $g$.
We pick up a complex structure on $\Sigma_g$.
We express $(z,\overline{z})$ a local coordinate, and denote it as $z=x+iy$ ($x,y\in\mathbb{R}$).
We define an almost complex structure $J:T_p\Sigma_g\to T_p\Sigma_g$ ($p\in \Sigma_g$) by
\begin{align}
J\left(\frac{\partial}{\partial x}\right):=\frac{\partial}{\partial y},\qquad
J\left(\frac{\partial}{\partial y}\right):=-\frac{\partial}{\partial x}.
\end{align}
This definition of $J$ is independent of the choice of local coordinates.
Since $J^2=-1$, eigenvalues of $J$ is $J=\pm i$.
We denote $T_p^{(1,0)}\Sigma_g$ and $T_p^{(0,1)}\Sigma_g$ the subspace of $T_p\Sigma_g$ whose eigenvalue of $J$ is $J=\pm i$ respectively, and denote $T_p^{\ast(1,0)}\Sigma_g$ and $T_p^{\ast(0,1)}\Sigma_g$ the corresponding cotangent bundles.
We define the globalization of $T_p^{\ast(1,0)}\Sigma_g$ and $T_p^{\ast(0,1)}\Sigma_g$, and we denote those bundles as $\Omega^1\Sigma_g$ and $\overline{\Omega}^1\Sigma_g$.
The cotangent bundle $\Omega^1\Sigma_g$ is a complex vector space of dimension $g$. We denote its basis as ${\omega_1, \cdots, \omega_g}$. The dual space of $\Omega^1\Sigma_g$ is denoted as $\check{\Omega}^1\Sigma_g$. The homology group $H_1(\Sigma_g, \mathbb{Z})$ on the Riemann surface $\Sigma_g$ can be regarded as a subspace of $\check{\Omega}^1\Sigma_g$ using the following map.
\begin{align}
H_1(\Sigma_g,\mathbb{Z})\to \check{\Omega}^1\Sigma_g,\qquad
[\gamma]\mapsto f_\gamma,\qquad
f_\gamma(\omega):=\int_\gamma\omega,\quad \omega\in \Omega^1\Sigma_g.
\label{dual-H1}
\end{align}
We can choose a symplectic basis of the homology group $H_1(\Sigma_g, \mathbb{Z})$ as $\alpha_j,\beta_j$ ($j=1,\cdots,g$).
The intersection number $I: H_1(\Sigma_g, \mathbb{Z}) \times H_1(\Sigma_g, \mathbb{Z}) \to \mathbb{Z}$ is given by 
\begin{align}
I(\alpha_i,\alpha_j)=I(\beta_i,\beta_j)=0,\qquad
I(\alpha_i,\beta_j)=\delta_{ij},\qquad \alpha_i,\beta_i\in H_1(\Sigma,\mathbb{Z}).\label{Sym-I}
\end{align}
We introduce the following vectors in $\mathbb{C}^g$:
\begin{align}
\Omega_j:=&\left(\int_{\alpha_j}\omega_1,\cdots,\int_{\alpha_j}\omega_g\right)\in\mathbb{C}^g,\quad
\Omega_{j+g}:=\left(\int_{\beta_j}\omega_1,\cdots,\int_{\beta_j}\omega_g\right)\in\mathbb{C}^g,\quad j=1,\cdots,g.
\label{syuuki-J}
\end{align}
We normalized the basis ${\omega_1, \cdots, \omega_g}$ of $\Omega^1\Sigma_g$ as
\begin{align}
\int_{\beta_i}\omega_j=\delta_{ij}.\label{tau-2}
\end{align}
The period matrix of the complex torus determined by the lattice spanned by (\ref{syuuki-J}) is a $2g\times g$ matrix whose elements are given by
\begin{align}
P(\Sigma_g):=\left(
\begin{array}{c}
\tau\\
I_g
\end{array}
\right),\quad
\tau:=\left(\int_{\alpha_i}\omega_j\right)_{i,j}.\label{R-Omega}
\end{align}
Here, $I_g$ is the $g\times g$ identity matrix.
This $\tau$ is valued in the Siegel upper plane ${\cal S}_g$:
\begin{align}
{\cal S}_g:=\{\tau\in{\rm Mat}(g,\mathbb{C})|\:\tau^T=\tau,\:{\rm Im}\tau>0\},\label{Sg}
\end{align}
where ${\rm Mat}(g,\mathbb{C})$ is the set of all $g\times g$ matrices with complex coefficients.
The complex lattice spanned by vectors (\ref{syuuki-J}) is
\begin{align}
{\rm Jac}(\tau):=\mathbb{C}^g/(\mathbb{Z}^g+\tau\cdot\mathbb{Z}^g),\label{Jacobi}
\end{align}
which is called the Jacobi variety.
We define a $\mathbb{Z}$-valued component $2g\times 2g$ matrix by
\begin{align}
E_J:=\left(
\begin{array}{cc}
0_g&1_g\\
-1_g&0_g
\end{array}
\right).\label{EJ}
\end{align}
This matrix satisfies ${}^TP(\Sigma_g)E_J^{-1}P(\Sigma_g)=E_J^{-1}$, and $\sqrt{-1}{}^TP(\Sigma_g)E_J^{-1}\overline{P}(\Sigma_g)>0$.
Therefore ${\rm Jac}(\tau)$ is a principal polarized Abelian variety.
See Appendix A for details of the Abelian variety.

The matrix $E_J$ is invariant under the action of the modular group ${\rm Sp}_{2g}(\mathbb{Z})$.
Consider an element of the modular group:
\begin{align}
T=\left(
\begin{array}{cc}
A&B\\
C&D
\end{array}
\right)\in {\rm Sp}_{2g}(\mathbb{Z}).\label{E-M}
\end{align}
Here, $A,B,C,D$ are $g\times g$ matrices.
From the definition of the matrix $E_J$, the modular transformation $T$ acts on the period matrix as $P(\Sigma_g)\mapsto TP(\Sigma_g)$ as the product of matrices.
By a basis change of the Jacobi variety ${\rm Jac}(\tau)$, the period matrix and matrix $E_J$ can be written as the Frobenius form in Appendix A.
The modular transformation $T$ acts on the basis $\omega_j$ (\ref{tau-2}), $\omega_j\mapsto T\cdot \omega_j$, which corresponds to a change of basis:
\begin{align}
\left(
\begin{array}{c}
\omega_1\\
\vdots\\
\omega_g
\end{array}
\right)
\mapsto
\left(
\begin{array}{c}
T\cdot \omega_1\\
\vdots\\
T\cdot \omega_g
\end{array}
\right)
=
 X\cdot
 \left(
\begin{array}{c}
\omega_1\\
\vdots\\
\omega_g
\end{array}
\right),\qquad X\in{\rm M}_g(\mathbb{C}),\quad {\rm det}X\neq 0.\label{X}
\end{align}
Here, we normalized $T\cdot\omega_j$ as
\begin{align}
\int_{T\cdot \beta_j}T\cdot \omega_k=\delta_{jk},\label{Mod-beta}
\end{align}
and the modular transformation $T$ acts on the $\beta_j$-cycles as $\beta_j\mapsto T\cdot \beta_j$. 
Therefore, the period matrix of the Jacobi variety undergoes the following transformation:
\begin{align}
\left(
\begin{array}{c}
\tau\\
1
\end{array}
\right)
\mapsto 
\left(
\begin{array}{cc}
A&B\\
C&D
\end{array}
\right)
\left(
\begin{array}{c}
\tau\\
1
\end{array}
\right)
X^T
=:\left(
\begin{array}{c}
\tilde{\tau}\\
1
\end{array}
\right).
\end{align}
Thus, the basis $\omega_j$ is transformed as
\begin{align}
(\omega_1, \cdots, \omega_g) \mapsto (\omega_1, \cdots, \omega_g) \cdot (C\tau+D)^{-1},\label{Mod-omega}
\end{align}
and the period matrix $\tau$ and the coordinate on the Jacobi variety $z\in\mathbb{C}^g/(\tau\cdot \mathbb{Z}^g+\mathbb{Z}^g)$ are transformed as
\begin{align}
\tau\mapsto\tilde{\tau}:=(A\tau+B)(C\tau+D)^{-1},\quad
z\mapsto \tilde{z}:=z\cdot((C\tau+D)^{-1})^T.\label{MOD-S}
\end{align}
These transformation are coincide with \cite{Ig64}.
The Jacobi variety as an Abelian variety is classified by the quotient space of the Siegel upper plane ${\cal S}_g$ with the modular group action ${\cal S}_g/{\rm Sp}_{2g}(\mathbb{Z})$, see Appendix A for details of the Abelian variety.

A mapping class group induces a symplectic automorphism ${\rm Sp}(H_1(\Sigma_g,\mathbb{Z})\otimes\mathbb{Z}_2)$ on the symplectic space $\{H_1(\Sigma_g,\mathbb{Z})\otimes\mathbb{Z}_2,I\}$,
 where $I$ is the intersection number (\ref{Sym-I}).
 We will compare the modular transformation ${\rm Sp}_{2g}(\mathbb{Z})$ and the symplectic automorphism ${\rm Sp}(H_1(\Sigma_g,\mathbb{Z})\otimes\mathbb{Z}_2)$.
 For this purpose, we will construct a symplectic space whose symplectic automorphism is the modular group ${\rm Sp}_{2g}(\mathbb{Z})$ based on \cite{GF12}.
We will show that this symplectic space is isomorphic to the symplectic space $\{H_1(\Sigma_g,\mathbb{Z})\otimes\mathbb{Z}_2,I\}$ in Appendix C.

Let us construct a symplectic space whose symplectic automorphism is the modular group ${\rm Sp}_{2g}(\mathbb{Z})$.
The Jacobi variety ${\rm Jac}(\tau)$ and the Picard manifold ${\rm Pic}^0(\Sigma_g)$ are isomorphic as schemes over $\mathbb{C}$ (Theorem 12.2.1 in \cite{Yanagida-lec}).
\begin{align}
{\rm Jac}(\tau)\simeq {\rm Pic}^0(\Sigma_g).\label{Alb=Pic}
\end{align}
Here, the Picard manifold is defined by ${\rm Pic}^0(\Sigma_g):=H^1(\Sigma_g,{\cal O}_{\Sigma_g})/H^1(\Sigma_{g,h},\mathbb{Z})\simeq\mathbb{C}^g/\mathbb{Z}^{2g}$ (Proposition 10.3.1 in \cite{Yanagida-lec}), where ${\cal O}_{\Sigma_g}$ denotes the sheaf of rings of a Riemann surface considered as a complex analytic space, and $\Sigma_{g,h}$ is the associated complex analytic space of the Riemann surface $\Sigma_g$ (Theorem 10.1.1 in \cite{Yanagida-lec}).
We define the Jacobian by the following equation \cite{GF12}:
\begin{align}
J_2(\Sigma_g):=\{\eta\in {\rm Pic}^0(\Sigma_g)|\:\eta^{\otimes 2}={\cal O}_{\Sigma_g}\}.\label{J2}
\end{align}
$J_2(\Sigma_g)$ is a $\mathbb{Z}_2$-vector space.
We denote the Weil pairing as
\begin{align}
\left<\quad,\quad\right>_W: J_2(\Sigma_g)\times J_2(\Sigma_g)\to\mathbb{Z}_2,\label{Weil-pair}
\end{align}
which is defined in \cite{GF12}.
The Jacobian $J_2(\Sigma_g)$ with the Weil pairing $\left<\quad,\quad\right>_W$ defines a symplectic space $(J_2(\Sigma_g),\left<\quad,\quad\right>_W)$.
According to proposition 3.5 of \cite{GB15}, the Weil pairing remains invariant under the modular transformations ${\rm Sp}_{2g}(\mathbb{Z})$ on the Jacobi variety.
Therefore, the modular group ${\rm Sp}_{2g}(\mathbb{Z})$ action can be regarded as symplectic automorphisms on $(J_2(\Sigma_g),\left<\quad,\quad\right>_W)$.
In Appendix C, we showed that this symplectic space $(J_2(\Sigma_g),\left<\quad,\quad\right>_W)$ is isomorphic to the symplectic space $(H_1(\Sigma_g,\mathbb{Z})\otimes\mathbb{Z}_2,I)$:
\begin{align}
\varphi_{\rm AJ}:\:(J_2(\Sigma_g),\left<\quad,\quad\right>_W)\xrightarrow{\simeq}(H_1(\Sigma_g,\mathbb{Z})\otimes\mathbb{Z}_2,I).
\label{AJ-sym}
\end{align}
Here, the isomorphism $\varphi_{\rm AJ}$ is defined by using the Abel-Jacobi map defined by (\ref{Tate=H1}) in Appendix C.

\subsection{Modular Transformation of Spin Structure}
Since the worldsheet of a superstring is a Riemann surface with a spin structure (spin curve), we need to consider the modular transformation of the spin structure.
We will explain the modular transformation of a spin structure.

To consider the modular transformation of a spin structure, we will first characterize the spin structure by using the holonomy of a spinor bundle.
Consider a spin curve $\Sigma_g$ with genus $g$.
Transition function of a spinor bundle on $\Sigma_g$ is valued in ${\rm Spin}(2)=U(1)$.
Consider a loop $\gamma$ on $\Sigma_g$.
Since $\gamma$ is a submanifold of $\Sigma_g$, we can obtain a spin structure on $\gamma$ by restricting a spin structure on $\Sigma_g$.
The transition function on a spin structure on $\gamma$ is valued in $\{\pm 1\}$, which is the holonomy of a spin structure on $\Sigma_g$ along $\gamma$.
Therefore, we obtain the holonomy of a spinor bundle on $\Sigma_g$ along a loop $\gamma$ valued in $\{\pm 1\}$.
The holonomy of a spinor bundle along two homotopy equivalent loops on $\Sigma_g$ are equivalent.
Then, the holonomy of a spinor bundle along elements of $H_1(\Sigma_g,\mathbb{Z})$ is well-defined.
We denote ${\rm hol}(\alpha_j)$ and ${\rm hol}(\beta_j)$ ($j=1,\cdots,g$) the holonomy of a spinor bundle along a loop $\alpha_j$ and $\beta_j$, respectively.
Here, we define $m\in {\rm Hom}_\mathbb{Z}(H_1(\Sigma_g,\mathbb{Z}),\mathbb{Z}_2)$ by
\begin{align}
m\left(\sum_ja_j\alpha_j+b_j\beta_j\right):=\sum_ja_j{\rm hol}(\alpha_j)+b_j{\rm hol}(\beta_j),\qquad a_j,b_j\in\mathbb{Z}.
\label{Char2}
\end{align}
We can identify the linear map $m\in {\rm Hom}_\mathbb{Z}(H_1(\Sigma_g,\mathbb{Z}),\mathbb{Z}_2)$ as an element of $\mathbb{Z}_2^{2g}$ by
\begin{align}
m:=(m(\alpha_1),m(\alpha_2),\cdots,m(\alpha_g),m(\beta_1),\cdots,m(\beta_g))\in\mathbb{Z}_2^{2g}.
\end{align}
We call this $m\in\mathbb{Z}_2^{2g}$ characteristic of a spinor bundle on a Riemann surface $\Sigma_g$.
The Brown quadratic form $q_m:H_1(\Sigma_g,\mathbb{Z})\to\mathbb{Z}_2$ corresponding to a characteristic $m\in\mathbb{Z}_2^{2g}$ is defined as follows \cite{Roh-the}:
\begin{align}
q_m\left(
\sum_jx_j\alpha_j+y_j\beta_j\right):=\sum_jx_jy_j+\sum_jx_jm(\alpha_j)+\sum_jy_jm(\beta_j)\:{\rm mod}2,\qquad x_j,y_j\in\mathbb{Z}.
\label{Spinq}
\end{align}
It satisfies
\begin{align}
q_m(x+y)=q_m(x)+q_m(y)+I(x,y)\:{\rm mod}2.\label{B-qI}
\end{align}
There exists the following isomorphism \cite{Roh-the}:
\begin{align}
H^1(\Sigma_g,\mathbb{Z}_2)\simeq {\rm Hom}_\mathbb{Z}(H_1(\Sigma_g,\mathbb{Z}),\mathbb{Z}_2),\qquad
w\mapsto q_{m_w}.\label{Poin+UC2}
\end{align}
Here, $m_w\in\mathbb{Z}_2^{2g}$ is a characteristic which corresponds to a spin structure $w\in H^1(\Sigma,\mathbb{Z}_2)$.
This means the classification of spin structure on a Riemann surface by $H^1(\Sigma,\mathbb{Z}_2)$ is equivalent to that of by the characteristic.
We obtain a symplectic space $\{H_1(\Sigma_g,\mathbb{Z})\otimes\mathbb{Z}_2,I,q_m\}$ from a spin curve, where $I$ is the intersection number (\ref{Sym-I}).
The symplectic automorphism ${\rm Sp}(H_1(\Sigma_g,\mathbb{Z})\otimes\mathbb{Z}_2)$ acts on the Brown quadratic form $q_m$ as the form (\ref{q-transf}).
The Brown Arf-invariant is given by \cite{Roh-the}:
\begin{align}
{\rm Arf}(q_m)=\frac{1}{2^{2g}}\sum_{a\in H_1(\Sigma,\mathbb{Z})}(-1)^{q_m(a)}=e(m)\in\mathbb{Z}_2.
\label{BArf}
\end{align}
Here, we denote $e(m)$ the parity of a characteristic $m$, which is given as
\begin{align}
e(m):=m(\alpha_1)m(\beta_1)+m(\alpha_2)m(\beta_2)+\cdots+m(\alpha_g)m(\beta_g)\in\mathbb{Z}_2.
\label{parity}
\end{align}
We call $m\in\mathbb{Z}_2^{2g}$ is even (odd) if ${\rm Arf}(q_m)=0$ (${\rm Arf}(q_m)=1$).
As explained in the Appendix B, the set of all even (odd) characteristic is the orbit of the ${\rm Sp}(H_1(\Sigma_g,\mathbb{Z})\otimes\mathbb{Z}_2)$ action.

To determine the modular transformation of the characteristic $m\in\mathbb{Z}_2^{2g}$, we will define a quadratic form on the symplectic space $(J_2(\Sigma_g),\left<\quad,\quad\right>_W)$.
The theta characteristic of a genus $g$ Riemann surface $\Sigma_g$ is defined as follows \cite{GF12}:
\begin{align}
{\rm Th}(\Sigma_g):=\{\theta\in{\rm Pic}^{g-1}(\Sigma_g)|\:\theta^{\otimes 2}=K_{\Sigma_g}\}.\label{Th1}
\end{align}
Here, $K_{\Sigma_g}$ is the canonical line bundle on $\Sigma_g$.
For any theta characteristic $\theta\in {\rm Th}(\Sigma_g)$, we define a quadratic form $q_\theta:J_2(\Sigma_g)\to \mathbb{Z}_2$ \cite{GF12}:
\begin{align}
q_\theta(\eta):=h^0(\Sigma_g,\eta\otimes\theta)+h^0(\Sigma_g,\theta)\:{\rm mod}\:2,\qquad \eta\in J_2(\Sigma_g),\qquad \theta\in{\rm Th}(\Sigma_g).
\label{q-theta}
\end{align}
It satisfies \cite{GF12}:
\begin{align}
q_\theta(\eta+\epsilon)=q_\theta(\eta)+q_\theta(\epsilon)+\left<\eta,\epsilon\right>_W,\qquad
\eta,\epsilon\in J_2(\Sigma_g).\label{q-theta2}
\end{align}
We denote the symplectic basis of the $\mathbb{Z}_2$-vector space $(J_2(\Sigma_g),\left<\quad,\quad\right>_W)$ as $(\eta_1,\cdots,\eta_g,\epsilon_1,\cdots,\epsilon_g)$.
 The Arf-invariant corresponding to the quadratic form $q_\theta$ is defined as
\begin{align}
{\rm Arf}(q_\theta):=\sum_{i=1}^gq_\theta(\eta_i)\cdot q_\theta(\epsilon_i)\in\mathbb{Z}_2.\label{theta-arf}
\end{align}
This definition is independent of the choice of symplectic basis.
For a symplectic vector space $(V,\left<\quad,\quad\right>)$ over $\mathbb{Z}_2$, where $V$ is an even-dimensional vector space and $\left<\quad,\quad\right>$ is a symplectic form, we denote $Q(V)$ the set of all quadratic forms on $V$ with respect to the symplectic form $\left<\quad,\quad\right>$.
The following group isomorphism exists \cite{GF12}:
\begin{align}
{\rm Th}(\Sigma_g)\xrightarrow{\simeq} Q(J_2(\Sigma_g),\left<\quad,\quad\right>_W),\qquad
\theta\mapsto q_\theta.
\label{cha-qf}
\end{align}

We will consider the relation between quadratic forms of the form $q_\theta$ ($\theta\in{\rm Th}(\Sigma_g)$) and the Brown quadratic forms $q_m$ ($m\in\mathbb{Z}_2^{2g}$).
Since the mapping (\ref{AJ-sym}) is an isomorphism, there exists a unique $\eta,\epsilon\in J_2(\Sigma)$ for any $\alpha,\beta\in H_1(\Sigma_g,\mathbb{Z})\otimes\mathbb{Z}_2$, such that $\varphi_{{\rm AJ}}(\alpha)=\eta$ and $\varphi_{{\rm AJ}}(\beta)=\epsilon$.
Since (\ref{AJ-sym}) is an isomorphism of symplectic spaces, i.e,
\begin{align}
I(\alpha,\beta)=\left<\varphi_{\rm AJ}(\alpha),\varphi_{\rm AJ}(\beta)\right>_W,
\end{align}
we obtain
\begin{align}
q_m(\varphi_{{\rm AJ}}^{-1}(\eta+\epsilon))=q_m(\varphi_{{\rm AJ}}^{-1}(\eta))+q_m(\varphi_{{\rm AJ}}^{-1}(\epsilon))+\left<\eta,\epsilon\right>_W,\qquad
\forall\eta,\epsilon\in J_2(\Sigma).
\end{align}
Therefore, $q_m\circ\varphi_{{\rm AJ}}^{-1}$ is a quadratic form on the symplectic space $(J_2(\Sigma_g),\left<\quad\quad\right>_W)$.
Since (\ref{cha-qf}) is bijective, there exists a unique $\theta[m]\in{\rm Th}(\Sigma_g)$ for any characteristic $m\in\mathbb{Z}_2^{2g}$ such that
\begin{align}
q_m\circ\varphi_{{\rm AJ}}^{-1}=q_{\theta[m]}.\label{m=theta}
\end{align}
It is obvious that the following map defined by (\ref{m=theta}) is injective.
\begin{align}
\varphi:\mathbb{Z}_2^{2g}\to{\rm Th}(\Sigma_g),\qquad m\mapsto \theta[m].\label{m=theta2}
\end{align}
As explained in \cite{GF12}, the number of quadratic forms on the symplectic space $(J_2(\Sigma_g),\left<\quad\quad\right>_W)$ is $2^{2g}$.
Since (\ref{cha-qf}) is bijective, the number of elements in ${\rm Th}(\Sigma_g)$ is $2^{2g}$. Hence, (\ref{m=theta2}) is bijective.
By definition, this bijection (\ref{m=theta2}) is consistent under the symplectic automorphisms of both symplectic spaces equipped with quadratic forms.
We obtain the following isomorphism of symplectic spaces with quadratic forms (see Appendix B for details of isomorphism of symplectic space equipped with a quadratic form):
\begin{align}
\{H_1(\Sigma_g,\mathbb{Z})\otimes\mathbb{Z}_2,I,q_m\}\simeq \{J_2(\Sigma_g),\left<\quad,\quad\right>_W\,q_{\theta[m]}\}.\label{AJmap-q}
\end{align}

When we perform a transformation $T\in{\rm Sp}(J_2(\Sigma))={\rm Sp}_{2g}(\mathbb{Z})$ on the symplectic space $(J_2(\Sigma),\left<\quad,\quad\right>W)$, the quadratic form $q_\theta$ is transformed according to (\ref{q-transf}) as
\begin{align}
q_\theta\mapsto q_{T\cdot\theta}(x):=q_\theta(T^{-1}(x)).
\end{align}
From (\ref{m=theta}), the corresponding $q_m$ is transformed as follows:
\begin{align}
q_m=q_{\theta[m]}\circ\varphi_{{\rm AJ}}\mapsto q_{T\cdot m}:=q_{\theta[m]}\circ T^{-1}\circ\varphi_{{\rm AJ}}.
\end{align}
Then, the modular transformation of the characteristic $m\mapsto T\cdot m$ is given by (see section 3.2 of \cite{Roh-the}):
\begin{align}
T\cdot m:=\left(
\begin{array}{cc}
D&-C\\
-B&A
\end{array}
\right)m+
\left(
\begin{array}{c}
(C{}^tD)_0\\
(A{}^tB)_0
\end{array}
\right)\:{\rm mod}\:2,\quad
T=\left(
\begin{array}{cc}
A&B\\
C&D
\end{array}
\right)\in{\rm Sp}_{2g}(\mathbb{Z}).\label{Ig-spin}
\end{align}
This result (\ref{Ig-spin}) determines the modular transformation of a spin structure on a spin curve.

\subsection{Characteristic Theta Function}
We will introduce the characteristic theta function as a section on a line bundle on the Jacobi variety \cite{C-Li}, and then we consider the modular transformation of the characteristic theta function.

We define the characteristic theta function $\Theta_m:{\cal S}_g\times\mathbb{C}^g\to\mathbb{C}$ \cite{Ig64}:
\begin{align}
\Theta_{m}(\tau,z)
:=\sum_{n\in\mathbb{Z}^g}\exp\left\{
\pi i{}^t(n+\frac{a}{2})\tau(n+\frac{a}{2})
+2i\pi{}^t(n+\frac{a}{2})(z+\frac{b}{2})
\right\},\quad
m=(a,b)\in\mathbb{R}^g\times\mathbb{R}^g.\label{RT-func}
\end{align}
Here, ${\cal S}_g$ denotes the Siegel upper half-space (\ref{Sg}).
Let us explain that the characteristic theta function is a section of a line bundle on the Jacobi variety.
For arbitrary $a,b\in\mathbb{R}^g$, consider a family of holomorphic function of the form:
\begin{align}
u^{a,b}_{\tau\cdot m+n}(z):=\exp(2\pi i({}^ta\cdot n-{}^tb\cdot m))\cdot
\exp(2\pi i(-\frac{1}{2}{}^tm\cdot \tau\cdot m-{}^tm\cdot z)),\qquad m,n\in\mathbb{Z}^g.
\end{align}
Here, $z\in{\rm Jac}(\tau)=\mathbb{C}^g/(\tau\cdot\mathbb{Z}^g+\mathbb{Z}^g)$.
This family of holomorphic function satisfies
\begin{align}
u^{a,b}_{\xi_1}(z)\cdot u^{a,b}_{\xi_2}(z+\xi_1)=u^{a,b}_{\xi_1+\xi_2}(z),\qquad \xi_1,\xi_2\in \tau\mathbb{Z}^g+\mathbb{Z}^g.\label{uab-1}
\end{align}
We define family of line bundles over the Jacobi variety by using the trivial line bundles $L:\mathbb{C}^g\times \mathbb{C}\to\mathbb{C}^g$ over $\mathbb{C}^g$ and $u^{ab}_{\xi}(z)$:
\begin{align}
{\cal L}_{ab}:\:(\mathbb{C}^g\times \mathbb{C})/\sim_{u^{ab}}\to {\rm Jac}(\tau)=\mathbb{C}^g/(\tau\cdot\mathbb{Z}^g+\mathbb{Z}^g).\label{uab-2}
\end{align}
Here, the equivalence relation $\sim_{u^{ab}}$ is defined as
\footnote{
This is an equivalence relation, because by the relation (\ref{uab-1}), we obtain $u_0^{ab}(z)=1$ and $u_\xi^{ab}(z)^{-1}=u_{-\xi}^{ab}(z+\xi)$, thus satisfying the axioms of an equivalence relation. 
}
\begin{align}
(z,t)\sim_{u^{ab}}(z',t')\in\mathbb{C}^g\times \mathbb{C}\:\Leftrightarrow\:
z-z'=\xi\in\tau\cdot \mathbb{Z}^g+\mathbb{Z}^g,\:t=t'u^{ab}_\xi(z).
\end{align}
The characteristic theta function $\Theta_{(a,b)}(z,\tau)$ is a section of this line bundle ${\cal L}_{ab}$.
Under a modular transformation $T\in{\rm Sp}_{2g}(\mathbb{Z})$ of the form (\ref{E-M}), the characteristic theta function $\Theta_m$ ($m\in\mathbb{Z}_2^{2g}$) is transformed as \cite{Ig64}:
\begin{align}
&\Theta_{T\cdot m}((A\tau+B)(C\tau+D)^{-1},z\cdot((C\tau+D)^{-1})^T)
\notag \\
=&{\cal X}(T)\exp(2\pi i\phi_m(T)))
{\rm det}(C\tau+D)^{1/2}\exp(\pi i\cdot {}^tz(C\tau+D)^{-1}C z)\Theta_m(\tau,z).
\label{Ig-theta}
\end{align}
Here, $T\cdot m\in\mathbb{Z}_2^{2g}$ is given by (\ref{Ig-spin}), and $\phi_m(T)$ is defined as
\begin{align}
\phi_m(T):=-\frac{1}{8}\cdot ({}^ta{}^tBDa+{}^tb{}^tACb-2{}^ta{}^tBCb-2{}^t(A^tB)_0(Da-Cb)).\label{Ig-mul}
\end{align}
Here, $m=(a,b)\in\mathbb{Z}_2^g\times\mathbb{Z}_2^g$.
The function $\chi(T)$ satisfies $\chi(T)^8=1$ for $T\in {\rm Sp}_{2g}(\mathbb{Z})$.
The explicit form of ${\cal X}(T)$ for all $T\in{\rm Sp}_{2g}(\mathbb{Z})$ is obtained in \cite{Sie44}.
The part of the transformation rule (\ref{Ig-theta}) that does not depend on the period matrix $\tau$ is called the theta multiplier denoted by
\begin{align}
\gamma_m(T):={\cal X}(T)\exp(2\pi i\phi_m(T))).
\label{Ig-tmul}
\end{align}
The theta multiplier does not give a group homomorphism from ${\rm Sp}_{2g}(\mathbb{Z})$ to $U(1)$. This arises from the issue of the sign of ${\rm det}(C\tau+D)^{1/2}$ in the formula (\ref{Ig-theta}) defining the theta multiplier.
Some papers consider the $\mathbb{Z}$-extension of the modular group called the extended mapping class group whose action on the characteristic theta function is group homomorphisms \cite{knot-theta}.
We will introduce the extended mapping class group later.

\section{Superstring on Super Riemann Surface}
In the NSR formalism, the superstring is a quantum field theory on a spin curve called a worldsheet.
The partition function of the superstring theory should be defined in the modular invariant way.
In the paper \cite{Hok-Pho}, the superstring partition function is calculated by considering the worldsheet as a super Riemann surface.
In this section, we will explain the relation between a super Riemann surface and a spin curve.
We will show that the modular invariance corresponds to super-diffeomorphism invariance in genus $g\leq 3$ case.
We will introduce the superstring chiral measure for genus $g\leq 2$ cases.

\subsection{Super Riemann Surface and Spin Curve}
In this subsection, we will explain a super Riemann surface and relation to a spin curve.

Let us start by explaining the super Riemann surface.
Let $\Sigma_g$ be a genus $g$ Riemann surface, and $\overline{\Sigma}_g:=(\Sigma_g,{\cal A})$ be a space endowed with a sheaf of $\mathbb{Z}_2$-graded commutative ring ${\cal A}={\cal A}_0\oplus{\cal A}_1$ over $\mathbb{C}$, where ${\cal A}_i{\cal A}_j\subset{\cal A}_{i+j}$. Let ${\cal N}\subset {\cal A}$ be an ideal sheaf consisting of nilpotent elements (since we are considering sheaves of commutative rings if a commutative ring is an integral domain, the nilpotent elements form an ideal).
$\overline{\Sigma}_g:=(\Sigma_g,{\cal A})$ is called a super Riemann surface if it satisfies the following conditions \cite{Smoduli}.
\begin{itemize}
\item[(1)]
$(\Sigma_g,{\cal A}/{\cal N})$ is a complex analytic space, and ${\rm rank}_\mathbb{C}({\cal N}/{\cal N}^2)=1$.
\item[(2)]
${\cal N}/{\cal N}^2$ is a free module sheaf over ${\cal A}/{\cal N}$, we have the following ring isomorphism:
\footnote{
A supermanifold of type $(m|1)$ is split (Corollary 2.2 in \cite{DW-SM}).
}
\begin{align}
{\cal A}\simeq\wedge_{({\cal A}/{\cal N})}({\cal N}/{\cal N}^2)\qquad{\rm (ring}\:{\rm isomorphism)}.
\label{split}
\end{align}
\item[(3)]
There exists a sub-bundle ${\cal D}$ of dimension $(0|1)$ of $T\overline{\Sigma}_g$, such that for there exists a section $D$ of ${\cal D}$ such that $D^2 \in T\Sigma_g$ that is not proportional to $D$. 
This ${\cal D}$ is called the superconformal structure.
This condition is equivalent to the existence of a $(0|1)$-dimensional sub-bundle ${\cal D}$ of $T\overline{\Sigma}_g$ satisfying the following short exact sequence.
\begin{align}
0\to{\cal D}\to T\overline{\Sigma}_g\to {\cal D}^2\to 0.\label{SC}
\end{align}
Here, we use the definition section 2.1 of \cite{W-SRie}.
\end{itemize}
For a super Riemann surface $\overline{\Sigma}_g$, the Riemann surface $\Sigma_g$ is called the body.
By choosing an appropriate open cover $\Sigma_g=\bigcup_\alpha U_\alpha$, the local coordinates of $\overline{\Sigma}_g=(\Sigma_g,{\cal A})$ can be chosen as $z\otimes {\cal A}/{\cal N}[\theta]$. Here, $z\in\mathbb{C}$ and $\theta$ are valued in complex Grassmann number which satisfies the following conditions:
\begin{align}
\theta^2=\overline{\theta}^2=0,\quad\theta\overline{\theta}+\overline{\theta}\theta=0.
\end{align}
The tangent space of a super Riemann surface $\overline{\Sigma}_g$ is defined by section 2 of \cite{DeWi-SM}.
We denote $E_M{}^A$ supervielbein, where the index $M$ runs $z,\overline{z},\theta,\overline{\theta}$, and the index $A$ runs the corresponding framed indices.
We could introduce superconnection and super curvature \cite{DeWi-SM}.
Since $\Sigma_g$ is not a Minkowski space, we have no super Poincar\'e symmetry, and no $R$-symmetry, see section 1.1 of \cite{QFT-math1}.
A map $f:(\Sigma,{\cal A})\to(\Sigma',{\cal B})$ between two super Riemann surfaces is a pair of a reduced map $f_{\rm red}:\Sigma\to \Sigma'$ and a homomorphism of $\mathbb{Z}_2$-graded sheaves $f^\ast:f_{\rm red}^{-1}({\cal B})\to{\cal A}$.
If a map  $f:(X,{\cal A})\to(Y,{\cal B})$ is an isomorphism of sheaves of rings and $f_{\rm red}:\Sigma\to \Sigma'$ is a diffeomorphism, it is called a super-diffeomorphism \cite{Smoduli}.
Super Weyl transformation is a transformation of the rescaling of the supermetric.

It is known that there is a one-to-one correspondence between super Riemann surfaces and spin curves \cite{Bat79}.
Consider a genus $g$ spin curve $\Sigma_g$.
By the Batchelor's theorem \cite{Bat79}, there is one-to-one correspondence between an isomorphism class of super Riemann surface
\footnote{
 Isomorphism of two super Riemann surfaces on the same body is defined in definition1.6 of \cite{Bat79}.
 }
 $\overline{\Sigma}_g$ whose body $\Sigma_g$, and a complex line bundle ${\cal E}(\overline{\Sigma}_g)$ on $\Sigma_g$ which satisfies
\footnote{
This correspondence does not include the information of the conformal structure on $M$.
}
\begin{align}
({\cal E}(\overline{\Sigma}_g))^2=K(\Sigma_g).\label{spin-SM}
\end{align}
Here, $K(\Sigma_g)$ is the canonical line bundle on $\Sigma_g$.
By the condition (\ref{spin-SM}), we realize that the line bundle ${\cal E}(\overline{\Sigma}_g)$ is a spinor bundle on $\Sigma_g$ which is classified as elements of $H^1(\Sigma_g,\mathbb{Z}_2)$.
Therefore, the superstring worldsheet as a spin curve can be identified as a super Riemann surface.

\subsection{Super-Diffeomorphism and Modular Invariance}
When we consider the superstring on a spin curve, we need to construct the partition function in a modular invariant way.
As mentioned above, we can regard the worldsheet as a super Riemann surface.
We will show that in genus $g\leq 3$ case, the modular invariance corresponds to super-diffeomorphism invariance.

Let us consider the modular invariance from the viewpoint of the super Riemann surface.
We have the universal family of super Riemann surfaces on the supermoduli space (quotient space of super Teichm\"uller space with the super mapping class group), whose each fiber is a super Riemann surface and all fibers are transformed each other by infinitesimal super-diffeomorphisms \cite{Smoduli}.
If the supermoduli space is projective, which means there exists a well-defined projection from the superrmoduli space to the spin moduli space, by projecting out the base manifold \cite{DW-SM} and by using the Batchelor's theorem \cite{Bat79}, we obtain the universal family of the spin curve over the spin moduli space \cite{Smoduli}.
The spin Teichm\"uller space is constructed as the set of spin structures and a point on the Teichm\"uller space \cite{Smoduli}.
There is the following injection called the Torelli map (Torelli's theorem), see definition 2.7 in \cite{Gr10}:
\begin{align}
{\rm Jac}:{\rm Teich}(\Sigma_g)/{\rm Mod}(\Sigma_g)\to{\cal S}_g/{\rm Sp}_{2g}(\mathbb{Z}),
\qquad
[\Sigma_g]\mapsto {\rm Jac}(\tau).
\label{Mod-Thich}
\end{align}
Here, $[\Sigma_g]$ is the isomorphism class of $\Sigma_g$, ${\rm Teich}(\Sigma_g)$ is the Teichm\"uller space, ${\rm Mod}(\Sigma_g)$ is the set of large diffeomorphisms called the mapping class group of $\Sigma_g$, ${\cal S}_g$ is the Siegel upper plane (\ref{Siegel}), and ${\rm Sp}_{2g}(\mathbb{Z})$ is the symplectic group (\ref{para-MOD}).
In genus $g\leq 3$ this injection is a bijection.
Therefore, we can identify a universal family of super Riemann surfaces on the supermoduli space, and a universal family of spin curves over ${\cal S}_g/{\rm Sp}_{2g}(\mathbb{Z})$ when genus $g\leq 3$.
This means the modular invariance corresponds to super-diffeomorphism invariance in genus $g\leq 3$ cases.
However, the Torelli map (\ref{Mod-Thich}) is not a bijective (Schottky problem) in genus $g\geq 4$ cases, and the supermoduli space is not projected in genus $g\geq 5$ \cite{DW-SM}. Thus, the modular invariance does not correspond to super-diffeomorphism invariance for genus $g\geq 4$.

\subsection{Genus $g\leq 2$ Superstring Chiral Measure}
Let us consider superstring theory on a genus $g\leq 2$ spin curve $\Sigma_g$ with the target space $\mathbb{R}^{10}$.
We have bosonic string $X^\mu$ and a fermionic string $\Psi^\mu$ (the $\mu=1,\cdots,10$ is the $\mathbb{R}^{10}$ directions), where $X:\Sigma_g\to \mathbb{R}^{10}$ is Grassmann even and $\Psi:\Sigma_g\to\mathbb{R}^{10}$ is Grassmann odd.
Superstring chiral measure should be defined in the modular invariant way.

We will explain how to obtain the superstring chiral measure on a super Riemann surface.
In the genus $g\leq 2$ case, the modular invariance is equivalent to super-diffeomorphism invariance.
The infinitesimal superconformal transformations, infinitesimal super-diffeomorphisms which preserve the superconformal structure, are fixed by the Wess-Zumino gauge condition, whose ghost field is the $bc$-ghost \cite{Hok-Pho}.
By gauge-fixing of the infinitesimal quasi-superconformal transformations \cite{Hok-Pho} (see also section 4 of \cite{DPho1}), infinitesimal super-diffeomorphisms which change the superconformal structure, we obtain the $\beta\gamma$-ghost.
By gauge-fixing infinitesimal super-diffeomorphisms, and integral out of the Grassmann odd part of the supermoduli space, we obtain the genus $g\leq 2$ superstring chiral measure.
However, it remains a problem in calculating the genus $g\geq 3$ superstring chiral measure.

In the genus $1$ case, the partition function on a spin curve $\Sigma$ with characteristic $(a,b)\in\mathbb{Z}_2^2$ is given by
\begin{align}
Z_{(a,b)}=\frac{|\Theta_{(a,b)}(\tau)|^8}{|\eta(\tau)|^8}.\label{g=1,spin}
\end{align}
Here, $\eta(\tau)$ is the Dedekind $\eta$-function.
The modular invariant partition functions are given by summing over spin sectors (GSO projection), whose coefficients on each spin sector are given by using the Arf-invariant \cite{StringW,GSO}.
Type $0$B and Type $0$A superstring partition function is $Z^{(n)}$ ($n=0,1$), where
\begin{align}
Z^{(n)}=\frac{1}{2}\sum_{(a,b)\in\mathbb{Z}_2^2}(-1)^{n{\rm Arf}(q_{(a,b)})}Z_{(a,b)}\overline{Z}_{(a,b)},\quad n=0,1.\label{0AB}
\end{align}
Type IIB and Type IIA superstring path integral is given by
\begin{align}
Z^{(n_L,n_R)}=\frac{1}{4}\left(
\sum_{(a,b)\in\mathbb{Z}_2^2}(-1)^{n_L{\rm Arf}(q_{(a,b)})}Z_{(a,b)}
\right)
\left(
\sum_{(a',b'))\in\mathbb{Z}_2^2}(-1)^{n_R{\rm Arf}(q_{(a',b')})}Z_{(a',b')}
\right).
\label{2AB}
\end{align}
Here, $(n_L,n_R)=(0,0),(1,1)$ are called Type IIB, and $(n_L,n_R)=(0,1),(1,0)$ are called Type IIA.
The path integral $Z^{(n_L,n_R)}$ for $(n_L,n_R)=(0,0),(1,1)$ are different, and $Z^{(n_L,n_R)}$ for $(n_L,n_R)=(0,1),(1,0)$ are different.
This difference comes from the parity anomaly of a Majorana-Weyl fermion \cite{GSO}. 

In the genus $2$ case, a ${\rm Sp}_4(\mathbb{Z})$ invariant partition function is obtained by summation of spin even (odd) sectors \cite{Hok-Pho,DPho1}:
\begin{align}
Z=\int({\rm det}\:{\rm Im}\tau)^{-5}d\mu(\tau)\wedge\overline{d\mu(\tau)}.\label{DP4-PF}
\end{align}
Here, the chiral superstring measure $d\mu(\tau)$ is firstly given by \cite{DPho1}, and it was rewritten later by \cite{Gr}:
\begin{align}
d\mu(\tau):=\frac{\Xi^{(2)}(\tau)}{16\pi^6\Psi_{10}(\tau)}\prod_{I\leq J}d\tau_{IJ},
\label{DPhoCM}
\end{align}
where
\begin{align}
\Xi^{(2)}:=&\frac{1}{4}(G_0^{(2)}-G_1^{(2)}+2G_2^{(2)}),
\notag \\
\Psi_{10}:=&\prod_\delta\Theta_\delta(\tau)^2,\quad
G_0^{(2)}:=\Theta_{(0,0,0,0)}(\tau)^{16},
\quad
G_1^{(2)}:=\Theta_{(0,0,0,0)}(\tau)^8\sum_{v\neq 0:\:{\rm even}}\Theta_v^8,
\notag \\
G_2^{(2)}:=&\Theta_{(0,0,0,0)}^4\sum_{v\neq w,\:v,w\neq 0:\:{\rm even}}\Theta_v(\tau)^4\Theta_w(\tau)^4\Theta_{v+w}(\tau)^4.
\label{Gi(2)}
\end{align}
The superstring chiral measure (\ref{DPhoCM}) is zero as in equation (1.8) in \cite{DPho4}, which means it is the superstring chiral measure of Type II.
At present, no other modular invariant superstring chiral measure is known for genus $2$.
Functions $G_i^{(2)}$ are modular forms of weight $8$ of the subgroup of the modular group $\Gamma_2(1,2)\subset {\rm Sp}_4(\mathbb{Z})$, where $\Gamma_2(1,2)$ is defined in \cite{Ig64}.
By theorem 1 of \cite{Ig-g2(2)}, any modular form of weight $8$ of $\Gamma_2(1,2)$ spanned by $16$th products of theta functions.
It is known the complex dimensions of vector space of the modular forms of weight $8$ of $\Gamma_2(1,2)$ is $4$ (\cite{Ig-g2(2)} in page p.405).
$G_i^{(2)}$ ($i=0,1,2$) cannot span the vector space of the modular forms of weight $8$ of $\Gamma_2(1,2)$.

To explain the meaning of $G_i^{(2)}$, we introduce a function on the Jacobi variety of genus $g$ Riemann surface \cite{Gr,Man85}:
\begin{align}
P_M(\tau):=\prod_{m\in M}\Theta_m(\tau),\qquad M\subset\mathbb{Z}_2^{2g}.
\end{align}
If $M$ contain odd characteristic, we obtain $P_M=0$.
We define a function \cite{Gr}:
\begin{align}
P_{i,s}^{(g)}(\tau):=\sum_{V\subset\mathbb{Z}_2^{2g};\:{\rm dim}V=i}P_V(\tau)^s.
\end{align}
The function $P_{i,s}^{(g)}$ is a modular form of weight $2^{i-1}s$ for the subgroup $\Gamma_g(1,2)$ \cite{Ig64}.
Since the theta multiplier (\ref{Ig-tmul}) is valued in $\mathbb{Z}_8$, any modular form of weight $8$ for the modular group ${\rm Sp}_{2g}(\mathbb{Z})$ should be written in terms of products of theta functions can be expanded by linear combination of $\{P_{i,s}^{(g)}\}_{2^{i-1}s=8}$.
We define a function $G_i^{(g)}$ \cite{Gr}:
\begin{align}
G_i^{(g)}:=P_{i,2^{4-i}}^{(g)}.\label{G(g)}
\end{align}
In $g=2$ case, (\ref{G(g)}) coincides with $G_i^{(2)}$ defined in (\ref{Gi(2)}).
We also define
\begin{align}
\Xi^{(g)}:=\frac{1}{2^g}\sum_{i=0}^g(-1)^i2^{\frac{i(i-1)}{2}}G_i^{(g)}.\label{Xi(g)}
\end{align}
It coincides with (\ref{Gi(2)}) in $g=2$ case.
It is a modular form of weight $8$ of $\Gamma_g(1,2)$, and its restriction to ${\cal S}_k\times{\cal S}_{g-k}$ is equal to $\Xi^{(k)}\cdot \Xi^{(g-k)}$.
Moreover, $\Xi^{(g)}$ is the unique linear combination of $G_i^{(g)}$ that restricts to the decomposable locus in this way \cite{Gr}.

\section{$3$-Dimensional Gelca-Hamilton TQFT}
It would be interesting to study modular property of the superstring chiral measure as path integral of a $3$-dimensional bulk modular invariant theory.
In this section, we will explain the extended $3$-dimensional manifold whose boundary is the Jacobi variety, and extended mapping class group based on \cite{knot-theta}.
Then, we will introduce the theta series as sections on a line bundle on the Jacobi variety by geometric quantization, and the representation of the extended mapping class group on the theta series.
Finally, we introduce the Gelca-Hamilton TQFT, which is one in the class of Atiyah's extended surface TQFTs.

\subsection{Extended Surface and $3$-Dimensional Extended Manifold}
In this subsection, we will explain that the Jacobi variety can be identified as a pair of a Riemann surface and the Lagrangian, called an extended surface.
The bulk of an extended surface is an extended $3$-dimensional manifold, which is triple of a $3$-dimensional manifold whose boundary is a Riemann surface, the Lagrangian subspace, and an integer.

To define a bulk $3$-dimensional manifold of the Jacobi variety, we will rewrite the Jacobi variety in another way.
When we construct the Jacobi variety from a Riemann surface $\Sigma_g$, we choose the complex structure on $\Sigma_g$ and half of the symplectic basis (\ref{Sym-I}) of $H_1(\Sigma_g,\mathbb{Z})$, with the condition eq.(\ref{tau-2}).
Here, the Jacobi variety is parametrized by $\tau$.
It is equivalent to choose a Lagrangian subspace of the symplectic manifold $\{H_1(\Sigma_g,\mathbb{R}),I\}$:
\begin{align}
\mathbf{L}(\tau):=\left\{\left.\sum_{j=1}^gb_j\beta_j\right|\:b_j\in\mathbb{R}\right\}.\label{Lag-H1}
\end{align}
The pair $(\Sigma_g,\mathbf{L}(\tau))$ of the Riemann surface $\Sigma$ and the Lagrangian $\mathbf{L}(\tau)$ is called an extended surface.
We can identify ${\rm Jac}(\tau)$ as an extended surface $(\Sigma_g,\mathbf{L}(\tau))$, where the Lagrangian subspace $\mathbf{L}(\tau)$ is given by (\ref{Lag-H1}).

Consider a $3$-dimensional compact $3$-dimensional oriented manifold $M$ with boundary $\partial M=\Sigma_g$, where $\Sigma_g$ is a genus $g$ Riemann surface.
Let $M^\dag$ be a compact oriented $3$-manifold with a framing which satisfies \cite{KW-EMCG}:
\begin{align}
-\partial M^\dag=\partial M=\Sigma_g.\label{Mdag}
\end{align}
We choose a diffeomorphism of their boundaries $f:\partial M\to -\partial M^\dag$.
We construct a $3$-dimensional closed manifold $M\cup_f M^\dag$ by connecting two boundaries $\partial M$ and $-\partial M^\dag$ by $f$.
There is ambiguity in choosing a framing of the tangent bundle on the $3$-dimensional manifold $M\cup_fM^\dag$ \cite{KW-EMCG}.
By the Lickorish-Wallace theorem, the choice of a framing of the tangent bundle on $M\cup_fM^\dag$ corresponds to a selection of a framed link in $S^3$ \cite{KW-EMCG}.
\footnote{
Let $N$ be a smooth, compact, oriented $3$-dimensional manifold.
A framed link in $N$ is a smooth embedding in the interior of $N$ of a disjoint union of finitely many annuli $S^1\times [0,1]$  (see section5.3.1 of \cite{knot-theta}).
}
Since every closed, orientable, $3$-dimensional manifold is the boundary of a smooth, orientable, connected $4$-dimensional manifold (see proposition6.4 in \cite{knot-theta}), we consider a $4$-dimensional manifold $W$ which satisfies
\begin{align}
\partial W=M\cup_fM^\dag.\label{3d-4d}
\end{align}
Choosing a $4$-dimensional manifold $W$ corresponds to selecting a framing of the tangent bundle on $M\cup_fM^\dag$ \cite{KW-EMCG}.
Therefore, the pair $(M\cup_fM^\dag,W)$ determines the framed $3$-dimensional manifold $(M,L)$, where $L$ is a framed link in $S^3$.
The intersection form
\begin{align}
I_4:\:H_2(W,\mathbb{R})/{\rm torsion}\times H_2(W,\mathbb{R})/{\rm torsion}\to\mathbb{Z},\quad
(x,y)\mapsto x\cdot y,\label{I4}
\end{align}
is acquired by applying Poincare duality to the cup product:
\begin{align}
\cup: H^2(W,\mathbb{R})/{\rm torsion}\times H^2(W,\mathbb{R})/{\rm torsion}\to H^4(W,\mathbb{R}).\label{cup}
\end{align}
We denote the signature of this pairing (\ref{I4}) by $\sigma(W)\in\mathbb{Z}$.
If $\partial W=\partial W'=M\cup_fM^\dag$, $W$ and $W'$ determine the same framing on $M\cup_fM^\dag$ \cite{KW-EMCG}.
Thus, we denote it as
\begin{align}
\sigma(L):=\sigma(W).\label{signature}
\end{align}
For any $n\in\mathbb{Z}$, there exists a framed link $L$ in $S^3$ which satisfies $n=\sigma(L)$, and this framed link correspond to a framing on $M\cup_fM^\dag$ \cite{KW-EMCG}.
From a diffeomorphism $f:\partial M\to-\partial M^\dag$, we obtain the inclusion $i:\partial M\to M^\dag$.
We obtain a Lagrangian subspace $\mathbf{L}$ of $H_1(\partial M,\mathbb{R})$:
\footnote{
The kernel of the map $H_1(\partial M)\to H_1(M)$ induced by the inclusion $i:\partial M\to M$ is a Lagrangian subspace of $H_1(\partial M)$ for any $3$-dimensional manifold \cite{KW-EMCG}.
}
\begin{align}
\mathbf{L}={\rm ker}\{i_\ast:H_1(\partial M,\mathbb{R})\to H_1(M^\dag,\mathbb{R})\}\subset H_1(\partial M,\mathbb{R}).\label{Lag-3d}
\end{align}
Here, the map $i_\ast:H_1(\partial M,\mathbb{R})\to H_1(M^\dag,\mathbb{R})$ is induced by the inclusion $i:\partial M\to M^\dag$.

We define an extended $3$-manifold as a triple $(M,\mathbf{L},n)$, where $M$ is a $3$-dimensional manifold whose boundary is a Riemann surface, $\mathbf{L}\subset H_1(\partial M,\mathbb{R})$ is a Lagrangian subgroup (\ref{Lag-3d}), and $n\in\mathbb{Z}$.
We define boundary of an extended $3$-manifold \cite{KW-EMCG}:
\begin{align}
\partial(M,\mathbf{L},n):=(\partial M,\mathbf{L}).\label{E3d}
\end{align}
Since the Lagrangian subspace (\ref{Lag-3d}) is determined by the choice of $M^\dag$, we identify $(\Sigma_g,\mathbf{L})$ and $(\Sigma_g,M^\dag)$ \cite{KW-EMCG}.
In the same way, the bulk $3$-dimensional extended manifold $(M,\mathbf{L},n)$ defined in (\ref{E3d}) can be identified as the triple $(M,M^\dag,W)$, where $W$ is defined by (\ref{3d-4d}), and we denote (\ref{E3d}) as follows \cite{KW-EMCG}:
\begin{align}
\partial (M,M^\dag,W)=(\Sigma_g,M^\dag).\label{E3d-2}
\end{align}
Since the Jacobi variety is identified as an extended surface, we can conclude that a bulk of the Jacobi variety is an extended $3$-dimensional manifold.

\subsection{Extended Mapping Class Group}
In this subsection, we will explain a $\mathbb{Z}$-extension of the mapping class group based on \cite{knot-theta}.
The extended mapping class group is defined as the set of pairs $(\varphi,n)$, where $\varphi$ is a diffeomorphism on a Riemann surface, and $n$ is an integer.
The origin of the $\mathbb{Z}$-extension comes from the framing of a bulk $3$-dimensional manifold whose boundary is a Jacobi variety.

We will use the viewpoint (\ref{E3d-2}) to define the extended mapping class group below.
Consider two extended surfaces $(\Sigma_g,M_1^\dag)$ and $(\Sigma_g,M_2^\dag)$, and consider a diffeomorphism $\varphi:\Sigma_g\to\Sigma_g$.
We construct a $3$-dimensional closed manifold:
\begin{align}
M:=(-M_1^\dag)\cup_{{\rm id}}I_\varphi\cup_{{\rm id}}(-M_2^\dag).
\end{align}
Here, $I_\varphi$ is the mapping cylinder of $\varphi:\Sigma_1\to -\Sigma_2$.
\footnote{
Let $X,Y$ be topological spaces and $\varphi:X\to Y$ be a continuous function.
Then, the mapping cylinder $I_\varphi$ of $\varphi$ is \cite{knot-theta}:
\begin{align}
I_\varphi=(X\times[0,1]\cup Y)/\sim_\varphi.\label{Map-cyl}
\end{align}
Here, $\sim_\varphi$ identifies $(x,1)$ with $\varphi(x)$ for every $x\in X$.
}
As we mentioned above eq.(\ref{signature}), there is ambiguity in choosing a framing on the $3$-dimensional manifold $M$.
The framing choice on $M$ corresponds to a framed link $L$ in $S^3$.
There exists a framed link $L$ in $S^3$ such that $\sigma(L)=n$ for any $n\in\mathbb{Z}$ \cite{KW-EMCG}.
We denote those set as
\begin{align}
(\varphi,n):(\Sigma_g,\mathbf{L}(M_1^\dag))\to (\Sigma_g,\mathbf{L}(M_2^\dag)).\label{EMOR}
\end{align}
Here, we denote $\mathbf{L}(M_j^\dag)$ ($j=1,2$) the Lagrangian subspace determined by $M_j^\dag$ ($j=1,2$).
The mapping (\ref{EMOR}) is called an extended diffeomorphism.
The extended diffeomorphism can be seen as a $4$-dimensional cobordism, see \cite{KW-EMCG} for details of this viewpoint.
We define the composition $(\varphi',n')\circ (\varphi,n)$ of two extended morphisms $(\varphi,n):(\Sigma_g,\mathbf{L}_1)\to(\Sigma_g,\mathbf{L}_2)$ and $(\varphi',n'):(\Sigma_g,\mathbf{L}_2)\to(\Sigma_g,\mathbf{L}_3)$, as connected sum of two $4$-dimensional manifolds $(\varphi,n)$ and $(\varphi',n')$ \cite{KW-EMCG}.
By Wall's theorem, the composition of those two extended diffeomorphisms is given by:
\begin{align}
(\varphi',n')\circ (\varphi,n):=(\varphi'\circ \varphi,n'+n+I_{\rm Mas}((\varphi'\circ \varphi)_\ast \mathbf{L}_1,(f')_\ast \mathbf{L}_2,\mathbf{L}_3)).\label{EMCG2-mul}
\end{align}
Here, $I_{\rm Mas}$ is the Maslov index \cite{knot-theta}.
We define the extended mapping class group as follows \cite{knot-theta}:
\begin{align}
\widetilde{{\rm MCG}}(g):=\{(\varphi,n)|\:\varphi:\Sigma_g\to\Sigma_g:\:{\rm diffeomorphism},\:n\in\mathbb{Z}\}.\label{EMCG2}
\end{align}
Here, the multiplication rule is given by (\ref{EMCG2-mul}).
Since the extended mapping class group acts on extended surfaces (Jacobi variety) as (\ref{EMOR}), we obtain the induced morphism of the characteristic theta functions.
Therefore, we need to compare the action of the extended mapping class group and modular group on the characteristic theta functions.

\subsection{Theta Series}
It is difficult to obtain the induced morphisms of the extended mapping class group on the characteristic theta functions directly.
Gelca and Hamilton obtain the extended mapping class group action on a line bundle of the Jacobi variety, which is obtained by the geometric quantization of the Jacobi variety \cite{knot-theta}.
In this subsection, we will consider the geometric quantization of the Jacobi variety, and obtain the Hilbert space of the sections of a line bundle on the Jacobi variety.
We will explain a representation of the extended mapping class group on the line bundle obtained from the geometric quantization.

We will start from considering the geometric quantization of the Jacobi variety.
Let $(\mathbf{x},\mathbf{y})$ be the coordinate on $\mathbb{R}^g\times\mathbb{R}^g$.
We choose a local coordinate on the Jacobi variety by using the projection \cite{knot-theta}:
\begin{align}
p:\:\mathbb{R}^g\times\mathbb{R}^g\to{\rm Jac}(\tau)=\mathbb{C}^g/(\tau\cdot\mathbb{Z}^g+\mathbb{Z}^g),\qquad (\mathbf{x},\mathbf{y})\mapsto [\mathbf{z}:=\mathbf{x}+\tau\cdot\mathbf{y}].\label{R2g-Jac}
\end{align}
Here, $\mathbf{z}:=\mathbf{x}+\tau\cdot\mathbf{y}\in\mathbb{C}^g$ and $[\cdots]$ is the identification class.
We choose a $2$-form:
\begin{align}
\omega=\sum_{j=1}^gdx_j\wedge dy_j
\in H^2({\rm Jac}(\tau),\mathbb{R}),\label{2form-J}
\end{align}
and the K\"ahler polarization:
\begin{align}
\mathbf{F}:=\left\{\frac{\partial}{\partial\overline{z}_1},\cdots,\frac{\partial}{\partial\overline{z}_g}\right\}.\label{K-pol}
\end{align}
Since we add no additional information for the Jacobi variety (or extended surface) to construct this polarized symplectic manifold $\{{\rm Jac}(\tau),\omega,\mathbf{F}\}$, we can identify $\{{\rm Jac}(\tau),\omega,\mathbf{F}\}$ as an extended surface $(\Sigma_g,\mathbf{L}(\tau))$ \cite{knot-theta}.
We define the Planck constant to satisfy the Weil quantization condition:
\begin{align}
\frac{1}{h}\omega=\frac{1}{2\pi\hbar}\omega\in H^2({\rm Jac}(\tau),\mathbb{Z}).\label{2form-J2}
\end{align}
The Weil quantization condition (\ref{2form-J2}) is a necessary and sufficient condition for the existence of a line bundle whose curvature is given by the $2$-form which satisfies the Weil condition.
Then, we obtain \cite{knot-theta}:
\begin{align}
h=\frac{1}{N},\qquad N:{\rm even}.
\end{align}
By conducting the geometric quantization of the classical data $({\rm Jac}(\tau),\omega, \mathbf{F})$, we obtain a line bundle ${\cal L}\to {\rm Jac}(\tau)$, whose curvature is given by the two form (\ref{2form-J2}).
The set of all sections of this line bundle is called the Hilbert space and denoted by ${\cal H}_{N,g}(\mathbf{L}(\tau))$.
The Hilbert space ${\cal H}_{N,g}(\mathbf{L}(\tau))$ is given as follows \cite{knot-theta}:
\begin{align}
{\cal H}_{N,g}(\mathbf{L}(\tau)):=\left\{\left.
\sum_{\mu\in\mathbb{Z}_N^g}a_\mu \theta_{N,\mu}(\tau,z)\right|\:a_\mu\in\mathbb{C}\right\}.\label{Thil}
\end{align}
Here, $\theta_{N,\mu}(\tau,z)$ is the theta series, which is an orthonormal basis of the Hilbert space:
\begin{align}
\theta_{N,\mathbf{\mu}}(\tau,z):=\sum_{\mathbf{n}\in\mathbb{Z}^g}
\exp\left\{
2\pi iN\left[
\frac{1}{2}\left(\frac{\mathbf{\mu}}{N}+\mathbf{n}\right)^T\tau\left(\frac{\mathbf{\mu}}{N}+\mathbf{n}\right)
+\left(\frac{\mathbf{\mu}}{N}+\mathbf{n}\right)^Tz\right]\right\},\quad
\mathbf{\mu}\in\mathbb{Z}_N^g.\label{C-theta}
\end{align}

We will next consider the representation of the extended mapping class group on the theta series.
For any extended mapping class group $(\varphi,n):(\Sigma_g,\mathbf{L})\to(\Sigma'_g,\mathbf{L}')$, we can define the following representation defined in section 7.5.3 of \cite{knot-theta}:
\begin{align}
V_N^{(g)}(\varphi,n):{\cal H}_{N,g}(\mathbf{L})\to{\cal H}_{N,g}(\mathbf{L}'),\quad
\psi\mapsto e^{\frac{\pi i}{4}(n-\sigma(L_{\varphi,\mathbf{L},\mathbf{L}'}))}\Omega(L_\varphi)\psi.
\label{Vng}
\end{align}
Here, $L_{\varphi,\mathbf{L},\mathbf{L}'}$ is the link obtained defined in section 7.5.3 of \cite{knot-theta}.
$L_\varphi$ is the framed link in $\Sigma_g\times[0,1]$, where $\varphi$ is obtained by performing surgery on the framed link $L_\varphi$ defined in section 7.4.1 of \cite{knot-theta}.
$\Omega(L_\varphi)$ is the coloring of the link $L_\varphi$ by $\Omega$ \cite{knot-theta}.
By using Theorem 5.7 (a) of \cite{knot-theta}, and section 6 of \cite{knot-theta}, the representation (\ref{Vng}) is determined as the product of the skein module.
On the other hand, by the modular transformation on the Jacobi variety, the theta series is transformed by the induced mapping.
In the next subsection, we will compare the representation of the extended mapping class group and the modular transformation.

\subsection{Modular Group and Extended Mapping Class Group}
We will compare the extended mapping class group and the modular group.
For this purpose, we will show that the theta series can be expanded as linear combinations of the characteristic theta function.
Then, we will compare the modular transformation and extended mapping class group, by acting those group on theta series.

To show that the theta series can be expanded as linear combinations of the characteristic theta functions, we will first introduce the $N$-th order theta function based on \cite{Fay-theta}.
A holomorphic function $f:\mathbb{C}^g\to\mathbb{C}$ on $\mathbb{C}^g$ which satisfies the following condition is called $N$-th order theta function with characteristic $(\delta,\epsilon)\in\mathbb{Z}_2^{2g}$ \cite{Fay-theta}:
\begin{align}
f(z_1,\cdots,z_j+2\pi i,\cdots,z_g)=&e^{2\pi i\delta_j}
f(z),
\notag \\
f(z_1+\tau_{j1},\cdots,z_g+\tau_{jg})=&e^{1/2N\tau_{jj}-Nz_j-2\pi i\epsilon_j}
f(z),\label{n-theta}
\end{align}
where $\kappa,\lambda\in\mathbb{Z}^g$ and $z\in\mathbb{C}^g$.
We denote $V_{(\delta,\epsilon)}^{(N)}(\tau,\Sigma_g)$ the set of $N$-th order theta functions with characteristic $(\delta,\epsilon)\in\mathbb{Z}^{2g}_2$.
We obtain
\begin{align}
V_{(0,0)}^{(N)}(\tau,\Sigma_g)={\cal H}_{N,g}(\mathbf{L}(\tau)).\label{C-n}
\end{align}
Here, the vector space ${\cal H}_{N,g}(\mathbf{L}(\tau))$ is defined in (\ref{Thil}).
For each characteristic $(\delta,\epsilon)\in\mathbb{Z}^{2g}_2$, the vector space $V_{(\delta,\epsilon)}^{(N)}(\tau,\Sigma_g)$ is a $N^g$-dimensional vector space, and it is spanned by the characteristic theta functions of the form $\{\Theta_{(\frac{2(\delta+\rho)}{N},\epsilon)}(Nz,N\tau)\}_{\rho\in\mathbb{Z}_N^g}$ \cite{Fay-theta}.
We obtain
\begin{align}
{\cal H}_{N,g}(\mathbf{L}(\tau))=\left\{\left.
\sum_{\rho\in(\mathbb{Z}_N)^g}\alpha_\rho\Theta_{(\frac{2\rho}{N},0)}(Nz,N\tau)\right|
\alpha_\rho\in\mathbb{C}
\right\},\qquad N\in 2\mathbb{Z}.\label{Hi-RT}
\end{align}
Therefore, the theta series $\{\theta_{N,\mu}(\tau,z)\}_{\mu\in\mathbb{Z}_N^g}$ can be expanded uniquely by characteristic theta functions of the form $\Theta_{(\frac{2\rho}{N},0)}(Nz,N\tau)$.
In particular, in the $N=2$ case, we obtain the following equation from the definition of the characteristic theta function:
\begin{align}
\Theta_{(\rho,0)}
(2z,2\tau)=&\sum_{m\in\mathbb{Z}^g}
\exp\left\{2\pi i(m+\frac{\rho}{2})^T\tau(m+\frac{\rho}{2})
+4\pi i(m+\frac{\rho}{2})z\right\}
\notag \\
=&\theta_{N=2,\rho}(\tau,z).\label{even-theta}
\end{align}

By a modular transformation of the form (\ref{E-M}), the characteristic theta function is transformed as (\ref{Ig-theta}).
The modular transformation of the theta series of the form $\{\theta_{N=2,\mu}\}_{\mu\in\mathbb{Z}_2^{2g}}$ which is consistent with the modular transformation (\ref{Ig-theta}) is
\begin{align}
\rho(T):\:{\cal H}_{N=2,g}(\mathbf{L}(\tau))&\to {\cal H}_{N=2,g}(\mathbf{L}(T\cdot \tau)),
\notag \\
\theta_{N=2,\mu}(\tau,z)
&\mapsto
{\cal X}(T)\exp(2\pi i\phi_{(\mu,0)}(T)))
{\rm det}(C\tau+D)^{1/2}
\notag \\
&\times
\exp(\pi i\cdot {}^tz(C\tau+D)^{-1}C z)
\theta_{N=2,\mu}(\tau,z).
\label{EMCG-MCG}
\end{align}
Here, we use (\ref{MOD-S}).
By definition, it is a linear isomorphic transformation.
The discrete Fourier transformation associated to the pair $(\mathbf{L},\mathbf{L}')$ \cite{knot-theta}:
\begin{align}
{\cal F}_{\mathbf{L},\mathbf{L}'}:{\cal H}_{N,g}(\mathbf{L})\to{\cal H}_{N,g}(\mathbf{L}'),\label{disctrete}
\end{align}
Combining the mapping (\ref{EMCG-MCG}) and ${\cal F}_{\mathbf{L},\mathbf{L}'}$, we define the discrete Fourier transform:
\begin{align}
{\cal F}_{\mathbf{L}(\tau)}(T):={\cal F}_{\mathbf{L}(T\cdot \tau),\mathbf{L}(\tau)}\circ \rho(T).\label{disc-2}
\end{align}
By theorem 7.9 and theorem 7.10 of \cite{knot-theta}, up to multiplication by an eighth root of unity, the discrete Fourier transform ${\cal F}_{\mathbf{L}(\tau)}$ is equivalent to the representation of the extended mapping class group (\ref{Vng}) of the form $V_{N=2}^{(g)}(\varphi,n):{\cal H}_{N=2,g}(\mathbf{L}(\tau))\to{\cal H}_{N=2,g}(\mathbf{L}(\tau))$.
Therefore, by the explanation in section 7.5 of \cite{knot-theta}, we can conclude that the representation of the mapping class group $\varphi$ (\ref{EMCG-MCG}) is equivalent to the representation of the extended mapping class group $V_{N=2}^g(\varphi,n)$ up to multiplication by an eighth root of unity for arbitrary $n\in\mathbb{Z}$.

\subsection{Gelca-Hamilton TQFT}
By the result \cite{knot-theta}, there is an extended TQFT whose path integral on a genus $g$ handlebody is the theta series.
This TQFT is called the Gelca-Hamilton TQFT.
We will briefly introduce the Gelca-Hamilton TQFT, more details are found in \cite{knot-theta}.

To introduce the Gelca-Hamilton TQFT as an extended Atiyah's TQFT, we need to introduce two categories:
\begin{itemize}
\item
The category of extended surface ${\cal C}_{\rm ES}$ (Category of classical data), whose objects are extended surface (Jacobi variety), and the set of morphisms are the extended mapping class group.
\item
$2$-dimensional category ${\cal C}_\Theta$ (Quantum System), whose objects are the Hilbert spaces $\{{\cal H}_{N,g}(\mathbf{L})\}$, and morphisms are representation of the extended mapping class group.
\end{itemize}
We define a mapping:
\begin{align}
V_N^{(g)}:{\rm Ob}({\cal C}_{\rm ES})\to{\rm Ob}({\cal C}_\Theta),\qquad
(\Sigma_g,\mathbf{L})\mapsto {\cal H}_{N,g}(\mathbf{L}).\label{Vng2}
\end{align}
Combining (\ref{Vng}) and (\ref{Vng2}), we obtain a functor between those two categories:
\begin{align}
V_N^{(g)}:{\cal C}_{\rm ES}\to{\cal C}_\Theta.\label{Functor}
\end{align}
When two extended surfaces $(\Sigma,\mathbf{L})$ and $(\Sigma',\mathbf{L}')$ are boundary of an extended $3$-dimensional manifold, we denote the bulk $3$-dimensional extended manifold as
\begin{align}
\partial(M,\partial_-M=\Sigma,\partial_+M=\Sigma',\mathbf{L},\mathbf{L}',L,n):=(\Sigma,\mathbf{L})\cup(\Sigma',\mathbf{L}').\label{Cob}
\end{align}
In this case, there exists a morphism \cite{knot-theta}:
\begin{align}
Z_N(M,\partial_-M=\Sigma,\partial_+M=\Sigma',\mathbf{L},\mathbf{L}',L,n):\:{\cal H}_{N,g}(\mathbf{L})\to{\cal H}_{N,g}(\mathbf{L}').\label{Zn}
\end{align}
We assume that this morphism $Z_N$ satisfies Atiyah's axiom for the extended surface written in section 7.5 of \cite{knot-theta}.
We also assume that the following conditions are satisfied \cite{knot-theta}:
\begin{itemize}
\item[(a)]
It satisfies
\begin{align}
Z_N(H_g,\O,\Sigma_g,\{0\},\mathbf{L},L,0):V(\O)\to{\cal H}_{N,g}(\mathbf{L}),\qquad 1\mapsto \theta^{\tau(\mathbf{L})}_{N,[L]}(z).\label{theta-3d}
\end{align}
Here, we denote $[L]$ the homology class of $L$ in $H_1(H_g,\mathbb{Z}_N^g)=\mathbb{Z}_N^g$, $H_g$ is the genus $g$ handlebody, and $V(\O)$ is the Hilbert space on $\O$.

\item[(b)]
For extended morphism $(\varphi,n):(\Sigma,\mathbf{L}_1)\to(\Sigma,\mathbf{L}_2)$, the $V_N^{(g)}(\varphi,n)$ and the morphism $Z_N$ are related by
\begin{align}
V_N^{(g)}(\varphi,n)=Z_N(I_\varphi,-\Sigma,\Sigma,\mathbf{L}_1,\mathbf{L}_2,\O,n).\label{V=Z}
\end{align}
Here, $I_\varphi$ is the mapping cylinder of $\varphi$.

\item[(c)]
There is a constant $\kappa\in\mathbb{C}$ such that for every framed link $L$ in $S^3$,
\begin{align}
Z_N(S^3,\O,\O,\{0\},\{0\},L,0)=\kappa\left<L\right>\in{\cal L}_N(S^3)=\mathbb{C}:\:V(\O)\to& V(\O),
\notag \\
 1\mapsto& \kappa\theta_{N,[L]}^{\tau(\O)}(z).
\end{align}
The value of $\kappa$ is uniquely determined by the above and is $\kappa=1/\sqrt{N}$.
\end{itemize}
$Z_N$ that satisfies those conditions exists uniquely, and it is called the Gelca-Hamilton TQFT.
Below, we call $Z_N$ the path integral of the Gelca-Hamilton TQFT.
Since any even characteristic theta function can be obtained by modular transformation of the theta series (\ref{even-theta}), we obtain that the genus $g\leq 2$ superstring chiral measure can be expanded as a linear combination of products of elements of the Hilbert space of the Gelca-Hamilton TQFT.

\section{Superstring Chiral Measure and the Gelca-Hamilton TQFT}
In this section, we investigate the superstring chiral measure as the path integral of the Gelca-Hamilton TQFT on a bulk $3$-dimensional manifold. 
We will rewrite the genus $g\leq 2$ superstring chiral measure by the theta series.
Then, we will obtain the genus $g\leq 2$ superstring chiral measure as the path integral of the Gelca-Hamilton TQFT on $3$-dimensional bulk extended manifolds.
We will show that the modular transformation of the genus $g\leq 2$ superstring chiral measure can be seen as the extended mapping class group action on the bulk $3$-dimensional manifolds.
Although there is ambiguity in choosing a $\mathbb{Z}$-extension of the modular group, the action of the extended mapping class group on the superstring chiral measure does not depend on the choice of the $\mathbb{Z}$-extension.

\subsection{Superstring Chiral Measure by Theta Series}
We will rewrite the genus $g\leq 2$ superstring chiral measure by using theta series.
Any even characteristic theta function $\Theta_{(\delta,\epsilon)}(z,\tau)$ can be obtained by modular transformation of a characteristic theta function $\Theta_{(\rho,0)}$ ($\rho\in\mathbb{Z}_2^g$).
Since the characteristic theta function of the form $\Theta_{(\rho,0)}$ is equivalent to the theta series (\ref{even-theta}), we can write any characteristic theta function by modular transformations of the theta series.

We will first consider the genus $g=2$ case.
The genus $2$ superstring measure (\ref{Gi(2)}) is expanded by even characteristic theta functions.
Even spin structures in genus $2$ case are as follows:
\begin{align}
v_1=&(0,0,0,0),\:v_2=(0,0,0,1),\:v_3=(0,0,1,0),\:v_4=(0,0,1,1),
\notag \\
v_5=&(0,1,0,0),\:v_6=(0,1,1,0),\:v_7=(1,0,0,0)\:v_8=(1,0,0,1),
\notag \\
v_9=&(1,1,0,0),\:v_{10}=(1,1,1,1).\label{Even-basis}
\end{align}
To rewrite the genus $2$ superstring chiral measure by the theta series by using (\ref{even-theta}), we need to obtain all even characteristics by modular transformations of even characteristics of the form $(\rho,0)\in\mathbb{Z}_2^{2g}$.
By the transformation rule (\ref{Ig-spin}), we obtain
\begin{align}
v_j=\left(
\begin{array}{cc}
\1_2&B_j\\
0&\1_2
\end{array}
\right)\cdot\left(
\begin{array}{c}
\epsilon_j\\
0
\end{array}
\right),\qquad j\in\{2,3,4,6,8,10\},\qquad\epsilon_j\in\mathbb{Z}_2^2,\label{theta-3d(2)}
\end{align}
where $\1_2:={\rm diag}(1,1)$ and $B_j$ and $\epsilon_j$ are given by
\begin{align}
B_2=&\left(
\begin{array}{cc}
0&b\\
b&1
\end{array}
\right),\quad
B_3=\left(
\begin{array}{cc}
1&b\\
b&0
\end{array}
\right),\quad
B_4=\left(
\begin{array}{cc}
1&b\\
b&1
\end{array}
\right),
\notag \\
B_6=&\left(
\begin{array}{cc}
b+1&b\\
b&c
\end{array}
\right),\quad
B_8=\left(
\begin{array}{cc}
a&b\\
b&b+1
\end{array}
\right),\quad
B_{10}=\left(
\begin{array}{cc}
a&1\\
1&c
\end{array}
\right),
\notag \\
\epsilon_2=&(0,0),\quad
\epsilon_3=(0,0),\quad
\epsilon_4=(0,0),\quad
\epsilon_6=(0,1),\quad
\epsilon_8=(0,1),\quad
\epsilon_{10}=(1,1).
\end{align}
We obtain
\begin{align}
\Theta_{v_j}(\tau,z)=\phi_j\theta_{N=2,\kappa_j}(\tau_j/2,z).\label{theS=Rthe(2)}
\end{align}
Here, we define
\begin{align}
\kappa_j=\left\{
\begin{array}{ccc}
\epsilon_j&\qquad&j=2,3,4,6,8,10,
\\
(0,0)&&j=1,\\
(0,1)&&j=5,\\
(1,0)&&j=7,\\
(1,1)&&j=9,
\end{array}
\right.
\end{align}
and
\begin{align}
\tau_j=\left\{
\begin{array}{ccc}
\tau&\qquad&j=1,5,7,9,\\
\tau-B_j&&j=2,3,4,6,8,10,
\end{array}
\right.
\end{align}
and constants
\begin{align}
\phi_j:=\left\{
\begin{array}{ccc}
1,&\qquad&j=1,5,7,9,\\
\chi\left(
\begin{array}{cc}
1&B_j\\
0&1
\end{array}
\right)
\exp\left\{
2\pi i\phi_{(\epsilon_j,0)}\left(
\begin{array}{cc}
1&B_j\\
0&1
\end{array}
\right)\right\}
&&j=2,3,4,6,8,10,
\end{array}
\right.
\end{align}
where $\chi$ and $\phi_{(\epsilon,\delta)}$ are defined below (\ref{Ig-theta}).
$\phi_j$ is valued in the $8$th root of the unity.
Then, $\Psi_{10}(\tau)$ defined in (\ref{Gi(2)}) can be written as
\begin{align}
\Psi_{10}(\tau)
=\prod_{j=1}^{10}\phi_j^2\theta_{n=2,\kappa_j}(\tau_j/2,z=0)^2.\label{Psi10-TS}
\end{align}
In the same way, we obtain
\begin{align}
G_0^{(2)}&=\phi_1^{16}\theta_{n=2,\kappa_1}(\tau_1/2,z=0)^{16},
\notag \\
G_1^{(2)}&=\phi_1^8\theta_{n=2,\kappa_1}(\tau_1/2,z=0)^8\sum_{j=2}^{10}\phi_j^8\theta_{n=2,\kappa_j}(\tau_j/2,z=0)^8,
\notag \\
G_2^{(2)}&=\phi_1^4\theta_{n=2,\kappa_1}(\tau_1/2,z=0)^4
\notag \\
\times&
\sum_{2\leq j< k,\:v_j+v_k:{\rm even}}
\left\{
\phi_j^4\phi_k^4\phi_{[v_j+v_k]}^4
\theta_{n=2,\kappa_j}(\tau_j/2,z=0)^4\theta_{n=2,\kappa_k}(\tau_k/2,z=0)^4
\right.
\notag \\
&\left.
\theta_{n=2,\kappa[v_j+v_k]}(\tau_{\kappa[v_j+v_k]}/2,z=0)^4
\right\}.
\label{Schi-TS}
\end{align}
Here, we define $\kappa[v_j+v_k]\in\mathbb{Z}_2^2$ which correspond to the even spin characteristic $v_j+v_k$ by (\ref{theS=Rthe(2)}).
Substituting (\ref{Psi10-TS}) and (\ref{Schi-TS}) into (\ref{DPhoCM}), we obtain the genus $2$ chiral superstring measure by using the theta series.

Let us consider the genus $1$ case.
Since the superstring path integral for each spin sector $(a,b)\in\mathbb{Z}_2^2$ is obtained (\ref{g=1,spin}), and it is zero for odd spin structure, Type $0$A, and Type $0$B superstring chiral measure are equivalent, and Type IIA and Type IIB superstring chiral measure are equivalent.
In the same way as genus $2$ case, we obtain the following relation between the characteristic theta functions and theta series:
\begin{align}
\Theta_{(0,0)}(\tau,z)=&\theta_{N=2,0}(\tau,z),
\quad
\Theta_{(1,0)}(\tau,z)=\theta_{N=2,1}(\tau,z),
\quad
\Theta_{(0,1)}(\tau,z)=\kappa\theta_{N=2,0}(-\tau,z),
\notag\\
\kappa:=&\chi\left(
\begin{array}{cc}
-1&0\\
0&1
\end{array}
\right)\exp\left(
2\pi i\phi_{(0,0)}\left(
\begin{array}{cc}
-1&0\\
0&1
\end{array}
\right)
\right).
\label{g0-2theta}
\end{align}
Here, $\chi$ and $\phi_{(0,0)}$ are defined below (\ref{Ig-theta}).
$\kappa$ is valued in the eighth root of the unity.
The Type $0$A and Type $0$B partition functions (\ref{0AB}) are expanded by the theta series as
\begin{align}
Z^{(0)}=Z^{(1)}=\frac{1}{2|\eta(\tau)|^{16}}\left\{
|\theta_{N=2,0}(\tau,0)|^{16}
+|\theta_{N=2,1}(\tau,0)|^{16}
+|\theta_{N=2,0}(-\tau,0)|^{16}
\right\}.\label{Type0}
\end{align}
Type IIA and Type IIB partition functions (\ref{2AB}) are expanded as
\begin{align}
Z^{(n_L,n_R)}=&\frac{1}{4|\eta(\tau)|^{16}}
\left(
\theta_{N=2,0}(\tau,0)^8
+\theta_{N=2,1}(\tau,0)^8
+\theta_{N=2,0}(-\tau,0)^8
\right)
\notag \\
\times&
\left(
\overline{\theta_{N=2,0}(\tau,0)}^8
+\overline{\theta_{N=2,1}(\tau,0)}^8
+\overline{\theta_{N=2,0}(-\tau,0)}^8
\right),\quad n_L,n_R=0,1.
\label{Type2}
\end{align}

Therefore, we obtain the genus $g\leq 2$ superstring chiral measure as a linear combination of products of the theta series, which are elements on the Hilbert space of the Gelca-Hamilton TQFT.

\subsection{Superstring in the Viewpoint of the Gelca-Hamilton TQFT}
We will formulate the genus $g\leq 2$ superstring chiral measure as a path integral of the Gelca-Hamilton TQFT on $3$-dimensional bulk extended manifolds.

We will first consider the genus $2$ superstring chiral measure.
By using the axiom (\ref{theta-3d}) of the Gelca-Hamilton TQFT, it can be expressed by $Z_2$ of the Gelca-Hamilton TQFT, just replace $\theta_{N=2,\mu}^\tau(z)$ to $Z_2(H_2,\O,\Sigma,\{0\},\mathbf{L}(\tau),L_\mu,0)\cdot 1$, where $L_\mu$ is the framed link in $S^3$ corresponds to $\mu\in\mathbb{Z}_N^g$, and $H_2$ is the genus $2$ handlebody:
\begin{align}
\Psi_{10}(\tau)
&=\prod_{j=1}^{10}\phi_j^2
\left.\left\{
Z_2(H_2,\O,\Sigma,\{0\},\mathbf{L}(\tau_j/2),L_{\kappa_j},0)\cdot 1
\right\}^2\right|_{z=0},
\notag \\
G_0^{(2)}&=\phi_1^{16}
\left.\left\{
Z_2(H_2,\O,\Sigma,\{0\},\mathbf{L}(\tau_1/2),L_{\kappa_1},0)\cdot 1
\right\}^{16}\right|_{z=0},
\notag \\
G_1^{(2)}&=\phi_1^8\left.\left\{
Z_2(H_2,\O,\Sigma,\{0\},\mathbf{L}(\tau_1/2),L_{\kappa_1},0)\cdot 1
\right\}^8\right|_{z=0}
\notag \\
&\times
\sum_{j=2}^{10}\phi_j^8
\left.\left\{
Z_2(H_2,\O,\Sigma,\mathbf{L}(\tau_j/2),L_{\kappa_j},0)\cdot 1
\right\}^8\right|_{z=0},
\notag 
\end{align}
\begin{align}
G_2^{(2)}&=\phi_1^4
\left.\left\{
Z_2(H_2,\O,\Sigma,\{0\},\mathbf{L}(\tau_1/2),L_{\kappa_1},0)\cdot 1
\right\}^4\right|_{z=0}
\notag \\
\times&
\sum_{2\leq j< k,\:v_j+v_k:{\rm even}}
\left[
\phi_j^4\phi_k^4\phi_{[v_j+v_k]}^4
\left.\left\{
Z_2(H_2,\O,\Sigma,\mathbf{L}(\tau_j/2),L_{\kappa_j},0)\cdot 0
\right\}^4\right|_{z=0}
\right.
\notag \\
&\quad
\left.\left\{
Z_2(H_2,\O,\Sigma,\{0\},\mathbf{L}(\tau_k/2),L_{\kappa_k},0)\cdot 1
\right\}^4\right|_{z=0}
\notag \\
&\quad
\left.
\left.\left\{
Z_2(H_2,\O,\Sigma,\mathbf{L}(\tau_{\kappa[v_j+v_k]/2}),L_{\kappa[v_j+v_k]},0)\cdot 1
\right\}^4\right|_{z=0}
\right].\label{SCM-GH}
\end{align}
As mentioned in the previous section, $Z_2$ of an extended $3$-dimensional manifold can be seen as the path integral over a $3$-dimensional extended manifold.
The genus $2$ superstring chiral measure is obtained by the path integral over some $3$-dimensional extended manifolds whose boundaries are the genus $2$ Riemann surface $\Sigma$.

Let us consider the modular transformation of the genus $2$ superstring chiral measure from the viewpoint of the Gelca-Hamilton TQFT.
Consider an element of the mapping class group $\varphi$.
To consider modular transformation in the Gelca-Hamilton TQFT, we need to consider $\mathbb{Z}$-extension of it.
Let us consider an element of the extended mapping class group $(\varphi,n)$, where $n\in\mathbb{Z}$.
We define the mapping class group action on the genus $2$ superstring chiral measure as extended morphism $(\varphi,n)$ action on each $3$-dimensional path integral term in the superstring chiral measure (\ref{SCM-GH}):
\begin{align}
&(H_2,\O,\Sigma,\{0\},\mathbf{L}(\tau_\kappa/2),L_\kappa,0)
\notag \\
&\mapsto (H_2,\O,\Sigma,\{0\},\mathbf{L}(\tau_\kappa/2),L_\kappa,0)\cup(I_\varphi,-\Sigma,\Sigma,\mathbf{L}(\tau_\kappa/2),\mathbf{L}(T_\varphi\cdot (\tau_\kappa/2)),\O,n).\label{Map-cyl}
\end{align}
Here, we denote $T_\varphi$ a modular transformation corresponds to $\varphi$.
As explained in the previous subsection, the extended mapping class group representation (\ref{Vng}) is equivalent to the representation of the mapping class group defined in (\ref{EMCG-MCG}) up to multiplication of the eighth root of the unity.
From the definition of the representation (\ref{Vng}), this multiplication of the eighth root of the unity is canceled totally on the superstring chiral measure (\ref{SCM-GH}).
Therefore, this definition of the modular transformation action rule on the superstring chiral measure (\ref{Map-cyl}) is consistent with the modular transformation of the characteristic theta functions (\ref{Ig-theta}).
It does not depend on the choice of the $\mathbb{Z}$-extension $n\in\mathbb{Z}$.

In the same way, the genus $1$ superstring partition function can be written as the path integral of the Gelca-Hamilton TQFT.
The numerator of Type $0$A and Type $0$B partition function (\ref{Type0}) can be written as
\begin{align}
&\left|
Z_2(H_1,\O,\Sigma,\{0\},\mathbf{L}(\tau),L_0,0)\cdot 1\right|^{16}
+
\left|
Z_2(H_1,\O,\Sigma,\{0\},\mathbf{L}(\tau),L_1,0)\cdot 1\right|^{16}
\notag \\
&+
\left|
Z_2(H_1,\O,\Sigma,\{0\},\mathbf{L}(-\tau),L_0,0)\cdot 1\right|^{16},\label{Type0-N}
\end{align}
where $L_\mu$ is the framed link in $S^3$ corresponds to $\mu\in\mathbb{Z}_2$, see the axiom (\ref{theta-3d}), $H_1$ is the genus $1$ handlebody, $\Sigma$ is a genus $1$ Riemann surface.
The numerator of Type IIA and Type IIB partition function (\ref{Type2}) can be written as
\begin{align}
&\left\{
\left(
Z_2(H_1,\O,\Sigma,\{0\},\mathbf{L}(\tau),L_0,0)\cdot 1\right)^8
+
\left(
Z_2(H_1,\O,\Sigma,\{0\},\mathbf{L}(\tau),L_1,0)\cdot 1\right)^8
\right.
\notag \\
&\left.
+
\left(
Z_2(H_1,\O,\Sigma,\{0\},\mathbf{L}(-\tau),L_0,0)\cdot 1\right)^8
\right\}
\notag \\
\times &
\left\{
\overline{\left(
Z_2(H_1,\O,\Sigma,\{0\},\mathbf{L}(\tau),L_0,0)\cdot 1\right)}^8
+
\overline{\left(
Z_2(H_1,\O,\Sigma,\{0\},\mathbf{L}(\tau),L_1,0)\cdot 1\right)}^8
\right.
\notag \\
&\left.
+
\overline{\left(
Z_2(H_1,\O,\Sigma,\{0\},\mathbf{L}(-\tau),L_0,0)\cdot 1\right)}^8
\right\}.\label{Type2-N}
\end{align}
In the same way as the genus $2$ case, we define the modular transformation $T_\varphi$ (it is induced by an element of the mapping class group $\varphi$) on those numerators of the path integrals of Type $0$ and Type II theory as extended morphism $(\varphi,n)$ action on each $3$-dimensional path integral term in (\ref{Type0-N}) and (\ref{Type2-N}):
\begin{align}
&(H_1,\O,\Sigma,\{0\},\mathbf{L}(\tau),L_\mu,0)
\notag \\
&\mapsto (H_1,\O,\Sigma,\{0\},\mathbf{L}(\tau),L_\mu,0)\cup(I_\varphi,-\Sigma,\Sigma,\mathbf{L}(\tau),\mathbf{L}(T_\varphi\cdot\tau),\O,n),\quad \mu=0,1,\label{Map-cyl-g1}
\end{align}
The ambiguity of choosing an integer $n\in\mathbb{Z}$ does not contribute to the transformation of (\ref{Type0-N}) and (\ref{Type2-N}).
This definition of the modular transformation action rule on the superstring chiral measure (\ref{Map-cyl-g1}) is consistent with the modular transformation of the characteristic theta functions (\ref{Ig-theta}).

We can conclude that the genus $g\leq 2$ superstring chiral measure can be interpreted as path integral of the Gelca-Hamilton TQFT on bulk $3$-dimensional bulk extended manifolds.
The modular transformation of the superstring chiral measure can be seen as extended mapping class group action on the bulk $3$-dimensional manifolds.

\section{Conclusion and Discussion}
We formulated the genus $g\leq 2$ superstring chiral measure as path integral of bulk $3$-dimensional the Gelca-Hamilton TQFT.
The genus $g\leq 2$ superstring chiral measure can be obtained as the path integral of the Gelca-Hamilton TQFT on $3$-dimensional bulk extended manifolds.
In this viewpoint, the modular transformation of the superstring chiral measure can be understood as the extended mapping class group action.
Although there is ambiguity to choose a $\mathbb{Z}$-extension, the choice of a $\mathbb{Z}$-extension does not contribute to the transformation of the superstring chiral measure.

When we formulated the genus $g\leq 2$ type II superstring chiral measure by path integral of $3$-dimensional bulk the Gelca-Hamilton TQFT, we only use the fact that the genus $g\leq 2$ superstring chiral measure is expressed by characteristic theta functions with even spin structure and the fact that it is weight $8$ modular form of $\Gamma_g(1,2)$.
Classification of the superstring chiral measures cannot discussed from the viewpoint of bulk $3$-dimensional the Gelca-Hamilton TQFT.
In the genus $1$ case, the superstring chiral measures are classified using the Arf-invariant \cite{GSO}, and it is also discussed in the viewpoint of triality \cite{Tong-Tur}.

It remains a problem to obtain the superstring chiral measure in genus $g\geq 3$.
As discussed in section $3$, in genus $g\leq 3$, the modular invariance corresponds to the super-diffeomorphism invariance of the superstring on a super Riemann surface.
However, modular invariance and super-diffeomorphism invariance do not correspond for genus $g\geq 4$.

\section*{Acknoledgements}
The author is grateful to Katsushi Ito for useful advice and discussions.
The author is grateful to Yuji Tachikawa for useful comments and discussions, Sergei Gukov and Yosuke Imamura for discussion.
The author is grateful to Denjoe O'Connor, Tony Dorlas, David E.Evans, Parmeswaran Nair, and Giandomenico Palumbo for listening to my paper and support.
The author thanks Dmitry Noshchenko for suggesting some pdfs, and Joshua Y. L. Jones to correct my English.
S.K is supported by a postdoctoral Scholarship in DIAS.

 \appendix
 
\section{Abelian Variety}
In the following, we will explain Abelian variety based on \cite{Yanagida-lec}.
A $g$-dimensional complex torus is a quotient space $T:=V/\Lambda$ of a $g$-dimensional $\mathbb{C}$-linear space $V$ by a lattice $\Lambda := \sum_{j=1}^{2g}\mathbb{Z}\lambda_j$ ($\lambda_j\in V$).
We say that a lattice $\Lambda := \sum_{j=1}^{2g}\mathbb{Z}\lambda_j$ is rank $2g$ when $\lambda_j$ ($j=1,\cdots,2g$) are independent as a $\mathbb{Z}$-module.
Given a basis $e_1,\cdots,e_g$ of the $g$-dimensional $\mathbb{C}$-vector space $V$, the generators of $\Lambda$ can be expressed as follows:
\begin{align}
\lambda_j=\sum_{j=1}^g a_{j,k}e_k,\qquad a_{j,k}\in\mathbb{C}.
\end{align}
We define the period matrix of the complex torus $T:=V/\Lambda$ as follows:
\begin{align}
P(\Lambda):=\left(
\begin{array}{ccc}
a_{1,1}&\cdots&a_{1,g}\\
\vdots&&\vdots\\
a_{j,1}&\cdots&a_{j,g}\\
\vdots&&\vdots\\
a_{2g,1}&\cdots&a_{2g,g}
\end{array}
\right).\label{Period-M}
\end{align}

Let $V/\Lambda$ be a complex torus, and consider a ring sheaf ${\cal O}$ on it, and assume that $(V/\Lambda,{\cal O})$ is a scheme.
As explained in sections 11.2 and section 9.1 of \cite{Yanagida-lec}, there exists a ring sheaf over a complex torus which constructs a scheme over $\mathbb{C}$. 
If there exists a scheme morphism $\varphi:(V/\Lambda,{\cal O})\to\mathbb{C}P^n$ that is an embedding as a scheme over $\mathbb{C}$, then the scheme $(V/\Lambda,{\cal O})$ along with the scheme morphism $\varphi$ is called an abelian variety.
The condition that a complex torus $V/\Lambda$ is an Abelian variety equivalent to the existence of a matrix $E$ satisfying the following three conditions concerning the period matrix $P(\Lambda)$ of the complex torus \cite{Yanagida-lec}:
\begin{itemize}
\item[(1)]
$E\in {\rm Mat}(2g,\mathbb{Z})$, ${}^TE=-E$.
\item[(2)]
${}^TP(\Lambda)E^{-1}P(\Lambda)=0$.
\item[(3)]
$\sqrt{-1}{}^TP(\Lambda) E^{-1}\overline{P(\Lambda)}>0$.
\end{itemize}
If we appropriately choose the basis ${\lambda_1, \cdots, \lambda_{2g}}$ of $\Lambda$, we can express the matrices $E$ and the period matrix $P(\Lambda)$ in the following form (Frobenius theorem \cite{Yanagida-lec}):
\begin{align}
E=\left(
\begin{array}{cc}
0&\Delta\\
-\Delta&0
\end{array}
\right),\qquad
\Delta={\rm diag}(d_1,\cdots,d_g),\: d_i\in\mathbb{Z}_{>0},\: d_1|d_2|\cdots|d_g.\label{Frob}
\end{align}
\begin{align}
P(\Lambda)=\left(
\begin{array}{c}
\tau\Delta^{-1}\\
I_g
\end{array}
\right).\label{11.3.2}
\end{align}
Here, $\tau$ is a $g \times g$ matrix with complex coefficients, satisfying $\tau^T = \tau$ and ${\rm Im}(\tau) > 0$.
The basis ${\lambda_1, \cdots, \lambda_{2g}}$ which provides this standard form for $E$ is called a symplectic basis.
We define the Siegel upper half space:
\begin{align}
{\cal S}_g:=\{\tau\in {\rm Mat}(g,\mathbb{C})|\:\tau^T=\tau,\:{\rm Im}\tau>0\}.\label{Siegel}
\end{align}
Here, ${\rm Mat}(g,\mathbb{C})$ denotes the set of all $g \times g$ matrices with complex coefficients.
The change of basis which preserve $E$ are matrices $M$ which satisfy $MEM^T = E$ ($M \in {\rm GL}(2g,\mathbb{Z})$). 
We denote ${\rm Sp}(\Delta,\mathbb{Z})$ the set of those basis transformations, which is called the para modular group:
\begin{align}
{\rm Sp}(\Delta,\mathbb{Z}):=\left\{M\in{\rm GL}(2g,\mathbb{Z})\left|\: M\left(
\begin{array}{cc}
0&\Delta\\
-\Delta&0
\end{array}
\right)
M^T=\left(
\begin{array}{cc}
0&\Delta\\
-\Delta&0
\end{array}
\right)
\right.
\right\}.
\label{para-MOD}
\end{align}
In particular, ${\rm Sp}_{2g}(\mathbb{Z}):={\rm Sp}(I_g,\mathbb{Z})$ is called the integral symplectic group or Siegel modular group.
By a transformation
\begin{align}
M=\left(
\begin{array}{cc}
A&B\\
C&D
\end{array}
\right)\in {\rm Sp}(\Delta,\mathbb{Z}),
\qquad
A,B,C,D\in {\rm Mat}(g,\mathbb{Z}).
\end{align}
the period matrix $P(\Lambda)$ (\ref{11.3.2}) transformed as follows \cite{Yanagida-lec}:
\begin{align}
P(\Lambda)=\left(
\begin{array}{c}
\tau\Delta^{-1}\\
I_g
\end{array}
\right)\mapsto 
\left(
\begin{array}{c}
(A\tau+B\Delta)(C\tau+D\Delta)^{-1}\\
I_g
\end{array}
\right).
\end{align}
This corresponds to the following action on the Siegel upper half space.
\begin{align}
\tau\mapsto \tau':=(A\tau+B\Delta)(C\tau+D\Delta)^{-1}\Delta.
\end{align}

The existence of an ample invertible sheaf on the scheme of a complex torus $(V/\Lambda,{\cal O})$ is necessary and sufficient for the complex torus $(V/\Lambda,{\cal O})$ to be an Abelian variety, where ${\cal O}$ is an appropriate ring sheaf over $V/\Lambda$ explained in the previous page.
By the Appell-Humbert theorem and Theorem 11.2.4 in \cite{Yanagida-lec}, there one to one correspondence between an ample line bundle over a complex torus $T=\mathbb{C}^g/\Lambda$, and the set of Hermite formula over $\mathbb{C}^g$ and half-character explained below:
\begin{itemize}
\item[(1)]
For a line bundle $L$ over $T=\mathbb{C}^g/\Lambda$, there uniquely exists a positive definite Hermitian form $H(z,w)$ on $\mathbb{C}^g$ and a map $\chi: \Lambda \to U(1)$ with the following properties.
\begin{itemize}
\item[(A1)]
Let $A(z, w)$ be the imaginary part of $H(z, w)$. Then, for any $\lambda, \lambda' \in \Lambda$, we have $A(\lambda, \lambda') \in \mathbb{Z}$.
\item[(A2)]
For any $\lambda, \lambda' \in \Lambda$, we have $\chi(\lambda + \lambda') = \chi(\lambda) \cdot \chi(\lambda') \cdot \exp(\sqrt{-1}\pi A(\lambda, \lambda'))$.
\end{itemize}
This map $\chi$ is called a half-character.

\item[(2)]
For a positive definite Hermitian form $H(z,w)$ and $\chi: \Lambda \to U(1)$ satisfying (A1) and (A2) above, we define
\begin{align}
j(\lambda,z):=\chi(\lambda)\exp(\pi H(z,\lambda)+\pi H(\lambda,\lambda)/2).
\end{align}
We define an action of $\lambda\in\Lambda$ on $\mathbb{C}^g \times \mathbb{C}$ via
\begin{align}
\lambda:\mathbb{C}^g\times\mathbb{C}\to\mathbb{C}^g\times\mathbb{C},\qquad
(z,\zeta)\mapsto (z+\lambda,j(\lambda,z)\zeta).
\end{align}
Then, 
\begin{align}
L(H,\chi):=(\mathbb{C}^g\times \mathbb{C})/\Lambda,\label{App-H}
\end{align}
is a line bundle over $T$, with $H$ and $\chi$ being the Hermitian form and half-character determined by the correspondence in (1).
\end{itemize}
We introduce the following equivalence relation on ample line bundles $(T, L)$ over the Abelian variety $T$ \cite{Yanagida-lec}:
\begin{itemize}
\item
Two ample line bundles $(T, L_1)$ and $(T, L_2)$ are equivalent, denoted as $(T, L_1) \sim (T, L_2)$, if there exists some $a \in T$ such that $t_a^*L_1 \simeq L_2$. Here, $t_a: T \to T$ represents the translation map $t \mapsto t + a$.
\end{itemize}
The condition $(T, L_1) \sim (T, L_2)$, is equivalent to the condition that the positive definite Hermitian formula is equivalent for those line bundles $L_1,L_2$.
Furthermore, when the period matrix of the torus is given by the Frobenius formula (\ref{Frob}), the positive definite Hermitian formula is of the form:
\begin{align}
H^{-1}=\Delta^{-1}({\rm Im}\tau)\Delta^{-1},\qquad 
\Delta={\rm diag}(d_1,\cdots,d_g),\: d_i\in\mathbb{Z}_{>0},\: d_1|d_2|\cdots|d_g.\label{Her-del}
\end{align}
The equivalence class whose Hermitian formula is given by (\ref{Her-del}) is called $\Delta$-type polarized Abelian variety.
The quotient space ${\cal S}_g / {\rm Sp}(\Delta,\mathbb{Z})$ corresponds bijectively to the equivalence classes of $\Delta$-type polarized Abelian varieties \cite{Yanagida-lec}.

\section{Quadratic form and the Arf-invariant}
To explain quadratic form, we need to consider symplectic space.
We say that two $2g$-dimensional $\mathbb{Z}_2$ symplectic spaces $(V_j,\left<\quad,\quad\right>_j)$ ($j=1,2$) are isomorphic if there exists a linear isomorphism $T:V_1\to V_2$ which satisfies
\begin{align}
\left<Tx,Ty\right>_2=\left<x,y\right>_1,\qquad x,y\in V_1.
\end{align}
In particular, we consider the case $V_1=V_2=V$, $\left<\quad,\quad\right>_1=\left<\quad,\quad\right>_2$.
We consider a symplectic basis $(\eta_1,\cdots,\eta_g\epsilon_1,\cdots,\epsilon_g)$:
\begin{align}
\left<\eta_i,\epsilon_j\right>=-\left<\epsilon_j,\eta_i\right>=\delta_{ij},\qquad
\left<\eta_i,\eta_j\right>=\left<\epsilon_i,\epsilon_j\right>=0.
\label{basis}
\end{align}
We denote an element of $V$ as $\eta=\sum_j(a_j\eta_j+b_j\epsilon_j)=:(a_1,\cdots,a_g,b_1,\cdots,b_g)$.
We obtain
\begin{align}
\left<\eta,\epsilon\right>=\eta^T\left(
\begin{array}{cc}
0&I_g\\
-I_g&0
\end{array}
\right)\epsilon.
\end{align}
We define the symplectic group ${\rm Sp}(V)$ of $V$ as 
\begin{align}
{\rm Sp}(V):=\{M:\:{\rm linear}\:{\rm isomorphism}|\:\left<M\eta,M\epsilon\right>=\left<\eta,\epsilon\right>\}.\label{SpV}
\end{align}
${\rm Sp}(V)$ is the set of isomorphisms on a symplectic vector space $(V,\left<\quad,\quad\right>)$.

Let $V$ be a $2g$-dimensional vector space over $\mathbb{Z}_2$, and let $\left<\quad,\quad\right>:V\times V\to\mathbb{Z}_2$ be a symplectic form.
We introduce the quadratic form $q$ associated with this symplectic form:
\begin{align}
q(x+y)=q(x)+q(y)+\left<x,y\right>,\qquad
x,y\in V.\label{q-sym}
\end{align}
A symplectic vector space $(V,\left<\quad,\quad\right>)$ is identified by its automorphisms $T\in{\rm Sp}(V)$.
When a symplectic space has a quadratic form $q$, it is necessary to extend the definition of automorphisms of the symplectic space to include the quadratic form.
When an automorphism $T\in{\rm Sp}(V)$ acts on the symplectic space $(V,\left<\quad,\quad\right>)$, the automorphism $T\in{\rm Sp}(V)$ action on the quadratic form is defined by
\begin{align}
(T\cdot q)(x):=q(T^{-1}(x)),\qquad x\in V.\label{q-transf}
\end{align}
This transformation rule (\ref{q-transf}) preserves the equation (\ref{q-sym}), and thus it is well-defined automorphism action on a quadratic form.

In a $\mathbb{Z}_2$-symplectic space $(V, \left<\cdot,\cdot\right>)$ with a symplectic basis $(\eta_1, \cdots, \eta_g, \epsilon_1, \cdots, \epsilon_g)$, the Arf-invariant corresponding to the quadratic form $q$ is defined as follows:
\begin{align}
{\rm Arf}(q):=\sum_{i=1}^gq(\eta_i)\cdot q(\epsilon_i)\in\mathbb{Z}_2.\label{Arf-def}
\end{align}
This definition is independent of the choice of symplectic basis. 
Let $Q(V)^+$ (or $Q(V)^-$) be the set of all quadratic forms on $V$ with respect to the symplectic form $\left<\cdot,\cdot\right>$, whose Arf-invariants are valued in $0$ (or $1$) \cite{GF12}:
\begin{align}
Q(V)^+:=&\{q\in Q(V,\left<\quad,\quad\right>)|\:{\rm Arf}(q)=0\},
\notag \\
Q(V)^-:=&\{q\in Q(V,\left<\quad,\quad\right>)|\:{\rm Arf}(q)=1\}.
\label{Arfpm}
\end{align}
The number of elements in $Q(V)^+$ is $2^{g-1}(2^g+1)$, and the number of elements in $Q(V)^-$ is $2^{g-1}(2^g-1)$ \cite{GF12}. 
The automorphism ${\rm Sp}(V)$ action on $Q(V)^\pm$ is closed. 
Furthermore, it is known that ${\rm Sp}(V)$ acts transitively on the orbits over $Q(V)^\pm$ \cite{GF12}. 
We obtain the following bijection:
\begin{align}
Q(V,\left<\quad,\quad\right>)/{\rm Sp}(V)\simeq \mathbb{Z}_2,\qquad
[q]\mapsto {\rm Arf}(q).
\label{Arf-qsym}
\end{align}
Here, we denote $[q]$ the equivalence class of a qudratic form $q$.

\section{$(J_2(\Sigma_g),\left<\quad,\quad\right>_W)\simeq (H_1(\Sigma_g,\mathbb{Z})\otimes\mathbb{Z}_2,I)$}
We will show that the symplectic space $(J_2(\Sigma_g),\left<\quad,\quad\right>_W)$ is isomorphic to the symplectic space $(H_1(\Sigma_g,\mathbb{Z})\otimes\mathbb{Z}_2,I)$.
We denote ${\rm Jac}(\tau)(\mathbb{C})$ the set of rational points of the Jacobi variety ${\rm Jac}(\tau)$ over $\mathbb{C}$.
 From the Mordell-Weil theorem, ${\rm Jac}(\tau)(\mathbb{C})$ is shown to be a finitely generated abelian group. 
 Since ${\rm Jac}(\tau)(\mathbb{C})$ is a subset of the divisor ${\rm Div}({\rm Jac}(\tau))$, $[n]p \in {\rm Div}({\rm Jac}(\tau))$ is defined. 
Here, $[n]p := p + \cdots + p$ denotes $p$ added $n$ times. 
The $n$-torsion group of the Jacobi variety ${\rm Jac}(\tau)$ is defined as follows \cite{Na-Sury}:
\begin{align}
A_\tau[n]:=\{p\in {\rm Jac}(\tau)[\mathbb{C}]|\:[n]p=O\}.
\end{align}
Here, $O\in {\rm Jac}(\tau)(\mathbb{C})$ is the base point \cite{Na-Sury}. 
The following group isomorphism holds \cite{GB15}:
\begin{align}
A_\tau[n]\simeq (\mathbb{Z}_n)^{2g},\qquad n\neq 0.\label{En}
\end{align}
In particular, when $n=2$, this isomorphism becomes \cite{GF12}:
\begin{align}
(\mathbb{Z}_2)^{2g}\simeq A_\tau[2],\qquad (\epsilon,\delta)\mapsto \frac{\tau\cdot\epsilon+\delta}{2},\qquad\epsilon,\delta\in\mathbb{Z}_2^g.\label{Tate2}
\end{align}
Let $l\in\mathbb{N}$ be a prime number, and consider the mapping:
\begin{align}
A_\tau[l^{n+1}]\xrightarrow{[l]}A_\tau[l^n],\qquad p\mapsto [l]p.
\end{align}
The $l$-adic Tate module of ${\rm Jac}(\tau)$ is defined by using this map \cite{GB15}:
\begin{align}
T_l({\rm Jac}(\tau)):={\rm lim}_{\leftarrow m}A_\tau[l^m].
\end{align}
From Proposition 2.2 of \cite{GB15}, $T_l({\rm Jac}(\tau))$ is a free $\mathbb{Z}_l$-module of rank $2g$.
As mentioned in Remark 2.3 of \cite{GB15}, there is a group isomorphism between the Tate module and the \'{e}tale homology group:
\begin{align}
T_l({\rm Jac}(\tau))\simeq H_1({\rm Jac}(\tau),\mathbb{Z})\otimes\mathbb{Z}_l.\label{Tate-h1}
\end{align}
\'Etale topology treats schemes categorically, and defines a topology on the category by introducing coverings \cite{Mun}.
There exist continuous maps between schemes equipped with the \'etale topology and the Zariski topology \cite{Mun}.
The Abel-Jacobi map (a continuous map) $\varphi: \Sigma_g \to {\rm Jac}(\tau)$ \cite{Yanagida-lec} induces a homomorphism on homology:
\begin{align}
\varphi_\ast:H_1(\Sigma_g,\mathbb{Z})\to H_1({\rm Jac}(\tau),\mathbb{Z}).\label{AJ-ind}
\end{align}
Since the Abel-Jacobi map $\varphi: \Sigma_g \to {\rm Jac}(\tau)$ is injective, $\varphi_\ast$ is an injective homomorphism by the axioms of homology \cite{AJ-lec}.
Therefore, combining (\ref{Tate-h1}) and (\ref{AJ-ind}), we obtain an injection:
\begin{align}
\varphi_{\rm AJ}:
H_1(\Sigma_g,\mathbb{Z})\otimes\mathbb{Z}_l\to H_1({\rm Jac}(\tau),\mathbb{Z})\otimes \mathbb{Z}_l\simeq T_l({\rm Jac}(\tau)).\label{Tate=H1}
\end{align}
From Proposition 2.2 of \cite{GB15}, the $l$-adic Tate module $T_l({\rm Jac}(\tau))$ is a free $\mathbb{Z}_l$-module of rank $2g$.
Since both sides are free $\mathbb{Z}_l$-modules of rank $2g$, this injective homomorphism is an isomorphism.
When applying (\ref{Tate=H1}) for $l=2$, the left-hand side with the intersection number (\ref{Sym-I}) define a symplectic vector space $(H_1(\Sigma_g,\mathbb{Z})\otimes\mathbb{Z}_2,I)$.
The Tate module on the right-hand side also becomes a symplectic vector space by introducing the Weil pairing of the Tate module defined in section 3 of \cite{GB15}.
By using \cite{GB15,Na-Sury}, Weil pairing of the Tate module is written as the form:
\footnote{
The Weil pairing of the Tate module is different from the Weil pairing defined in (\ref{Weil-pair}).
}
\begin{align}
e_l:T_2({\rm Jac}(\tau))\times T_2({\rm Jac}(\tau))\to\mathbb{Z}_2.\label{TW}
\end{align}
By the isomorphism (\ref{Tate=H1}) and Proposition 3.5 of \cite{GB15}, a symplectic basis of the symplectic space $(H_1(\Sigma_g,\mathbb{Z})\otimes\mathbb{Z}_2,I)$ are mapped to a symplectic basis $\{\eta_j,\epsilon_j\}_{j=1,\cdots,g}$ of the Tate module $T_2({\rm Jac}(\tau))$:
\begin{align}
e_2(\eta_i,\eta_j)=e_2(\epsilon_i,\epsilon_j)=0,\qquad
e_2(\eta_i,\epsilon_j)=-e_2(\epsilon_j,\eta_i)=\delta_{ij}.
\label{W-basis}
\end{align}
Therefore, (\ref{Tate=H1}) gives an isomorphism between the Tate module $(T_2({\rm Jac}(\tau)),e_2)$ and $(H_1(\Sigma,\mathbb{Z})\otimes\mathbb{Z}_2,I)$. 
Furthermore, by using the Remark 3.2 of \cite{GB15} and \cite{PB}, we can conclude that the symplectic spaces $(T_2({\rm Jac}(\tau)),e_2)$ and $(J_2(\Sigma),\left<\quad,\quad\right>_W)$ (\ref{Weil-pair}) are isomorphic. 
Thus, $(H_1(\Sigma,\mathbb{Z})\otimes\mathbb{Z}_2,I)$ and $(J_2(\Sigma),\left<\quad,\quad\right>_W)$ are isomorphic as symplectic spaces.

\end{document}